\documentclass[11pt,a4paper]{article}

\pdfoutput = 1

\usepackage{jcappub}
\usepackage{multirow}
\pdfoutput=1
\usepackage[utf8]{inputenc}
\usepackage{float}
\usepackage{hyperref}
\usepackage{amsmath}
\usepackage{graphicx}
\usepackage{dcolumn}
\usepackage{bm}
\usepackage{epsfig}
\usepackage{amssymb,latexsym,mathrsfs}
\usepackage{graphicx}
\usepackage{color}
\usepackage{booktabs}

\usepackage{comment}

\definecolor{myred}{RGB}{180,50,28}
\definecolor{myblue}{RGB}{2,50,180}
\definecolor{mygreen}{RGB}{2,150,80}

\hypersetup{colorlinks=true,
            citecolor=blue,
            linkcolor=blue,
            urlcolor=blue}

\newcommand{\be}{\begin{equation}}
\newcommand{\ee}{\end{equation}}
\newcommand{\bea}{\begin{eqnarray}}
\newcommand{\eea}{\end{eqnarray}}

\newcommand{\mpl}{M_{*}^{2}}
\newcommand{\bra}{\alpha_{\textrm{B}}}
\newcommand{\run}{\alpha_{\textrm{M}}}
\newcommand{\kin}{\alpha_{\textrm{K}}}
\newcommand{\ten}{\alpha_{\textrm{T}}}

\newcommand{\csN}{c_{\mathrm{sN}}^{2}}
\newcommand{\dd}{{\rm{d}}}

\newcommand{\hiclass}{{\tt hi\_class}}

\renewcommand{\vec}[1]{\boldsymbol{#1}}

\newcommand{\barkap}{\Bar{\kappa}}
\newcommand{\Geff}{G_{\rm eff}}
\newcommand{\deltaG}{\frac{\delta G}{G}}
\newcommand{\vecq}{\vec{q}}
\newcommand{\vecx}{\vec{x}}
\newcommand{\xxi}{x_{i}}

\newcommand{\xxii}{x_{i,i}}
\newcommand{\Sii}{S_{i,i}}
\newcommand{\Sij}{S_{i,j}}
\newcommand{\Sji}{S_{j,i}}

\newcommand{\veck}{\vec{k}}

\title{Revisiting Vainshtein Screening for fast {\it N}-body simulations}

\author[a,1]{Guilherme Brando,\note{Corresponding author.}}
\author[b]{Kazuya Koyama,}
\author[c]{Hans A. Winther}

\affiliation[a]{Max Planck Institute for Gravitational Physics (Albert Einstein Institute) \\
Am Mühlenberg 1, 14476 Potsdam-Golm, Germany}
\affiliation[b]{Institute of Cosmology and Gravitation, University of Portsmouth,\\ Dennis Sciama Building, Burnaby Road, Portsmouth PO1 3FX, United Kingdom}
\affiliation[c]{Institute of Theoretical Astrophysics, University of Oslo,\\ PO Box 1029, Blindern 0315, Oslo,
Norway}

\emailAdd{guilherme.brando@aei.mpg.de}
\emailAdd{kazuya.koyama@port.ac.uk}
\emailAdd{h.a.winther@astro.uio.no}

\abstract{We revisit a method to incorporate the Vainshtein screening mechanism in {\it N}-body simulations proposed by R. Scoccimarro in~\cite{Scoccimarro:2009eu}. We further extend this method to cover a subset of Horndeski theories that evade the bound on the speed of gravitational waves set by the binary neutron star merger GW170817. The procedure consists of the computation of an effective gravitational coupling that is time \emph{and} scale dependent, $G_{\rm eff}\left(k,z\right)$, where the scale dependence will incorporate the screening of the fifth-force. This is a fast procedure that when contrasted to the alternative of solving the full equation of motion for the scalar field inside {\it N}-body codes, reduces considerably the computational time and complexity required to run simulations. To test the validity of this approach in the non-linear regime, we have implemented it in a COmoving Lagrangian Approximation (COLA) {\it N}-body code, and ran simulations for two gravity models that have full {\it N}-body simulation outputs available in the literature, nDGP and Cubic Galileon. We validate the combination of the COLA method with this implementation of the Vainshtein mechanism with full {\it N}-body simulations for predicting the boost function: the ratio between the modified gravity non-linear matter power spectrum and its General Relativity counterpart. This quantity is of great importance for building emulators in beyond-$\Lambda$CDM models, and we find that the method described in this work has an agreement of below $2\%$ for scales down to $k \approx 3h/$Mpc with respect to full {\it N}-body simulations.}
\begin{document}
\maketitle
\flushbottom

\section{Introduction} \label{intro}

The Large Scale Structure (LSS) of the Universe is known to contain vast wealth of information regarding how the initially linear and Gaussian perturbations evolved into the highly non-linear, non-Gaussian and complex structure we observe in the sky. Ongoing, DESI~\cite{DESI:2016fyo}, and upcoming galaxy surveys, Euclid~\cite{EUCLID:2011zbd} and Rubin observatory's LSST~\cite{LSSTScience:2009jmu}, will be able to give us information on how this evolution took place with unprecedented precision. During much of the past decade, the effort of reducing instrumental and systematical errors of galaxy surveys, in order to increase the constraining power on cosmological parameters, has occupied one of the central topics in LSS cosmology. Notwithstanding, an analogous task of increasing the accuracy on the theoretical modelling of the Universe in this highly non-linear regime has also been pursued by many. In order to probe these scales, the main focus on exploring the non-linear nature of the Universe is on the matter two-point correlation function and its Fourier transform, the matter power spectrum. To accurately compute this quantity in the deeply non-linear regime, a common approach is to use {\it N}-body simulations that evolve the cold dark matter (CDM) particles by solving the geodesic and Poisson equations following a given theory of gravity, which within the Standard Model of Cosmology, $\Lambda$CDM, is given by Einstein's General Relativity (GR).

However, the accurate data coming from the above mentioned surveys opens up the possibility of probing the nature of dark energy, the elusive and exotic fluid of negative pressure responsible for the current acceleration of our Universe. In GR the simplest solution to this fluid is to add a constant term in the Einstein field equations, the so-called Cosmological Constant $\Lambda$. Nevertheless, this simple solution faces conceptual and observational problems, which, in turn, prompted cosmologists to look for alternatives to this paradigm. Even though at small and dense regions, GR is known to correctly describe gravity due to the tight constraints from Solar-System and astrophysical tests, tests of gravity on the largest scales of our Universe are still not constraining enough. Therefore, in order to accommodate the late-time acceleration of our Universe, one possibility is to modify Einstein's theory of gravity on these large scales. Usually, modified theories of gravity (MG) add an extra degree of freedom, and the simplest MG theories of gravity are scalar-tensor theories, in which an extra scalar field is added to the Einstein-Hilbert action and coupled to the matter fields. One of the main new features of these theories is the emergence of a new force, called a fifth-force, that will act on test particles. So far, much of the exploration of these theories has been restricted to tests that probe directly the cosmological background or that probe the linear scales of the LSS of our Universe, as, in both regimes, we can quickly generate observables using Einstein-Boltzmann solvers for MG theories. In turn, this allows us to run full Monte Carlo Markov Chains (MCMC) to produce constraints on cosmological and MG parameters. Despite the simplicity on modelling linear scales, the impact MG leaves on the structure formation in our Universe is distinctive enough to push us to look for different alternatives to model the non-linear matter power spectrum.

Full {\it N}-body simulations are known to be time consuming and computationally expensive, limiting our ability to use them as a quick and cheap method to generate non-linear matter power spectra. The cost of running {\it N}-body simulations in MG theories is even worse, as one needs to solve an extra equation of motion for the scalar field fluctuations.  This equation is a non-linear equation, which adds another layer of complexity, as one often needs to solve it using a different scheme than the one implemented to solve the Poisson equation. Additionally, these theories are endowed with screening mechanisms, mechanisms that are capable of suppressing this extra force, guaranteeing that in dense environments, MG theories recover GR predictions, thus, satisfying Solar-System and laboratory tests of gravity~\cite{Koyama:2015vza,Clifton:2011jh,Ferreira:2019xrr}.

Current MG simulations~\cite{Arnold:2019vpg,Ruan:2021wup} usually take from a factor of $2$ up to a factor of $10$ longer than a $\Lambda$CDM one, and performing a full MCMC parameter estimation with simulations simply becomes not a feasible task, as one would need an order of $10^{4}$ simulations, or even more, to successfully sample the parameter space and reach convergence on the chains. Therefore, emulation techniques have been put forward in the community as viable alternatives to bypass this issue in standard cosmologies~\cite{Heitmann:2006hr,Habib:2007ca,Heitmann:2008eq,Heitmann:2009cu,Lawrence:2009uk,Agarwal:2012ew,Heitmann:2015xma,Lawrence:2017ost,Bocquet:2020tes,Kwan:2012nd,DeRose:2018xdj,McClintock:2018uyf,Zhai:2018plk,Nishimichi:2018etk,Kobayashi:2020zsw,Miyatake:2020uhg,Cuesta-Lazaro:2022dgr,Donald-McCann:2021nxc,Maksimova:2021ynf,Yuan:2022jqf,Euclid:2020rfv,Angulo:2020vky}, as well as in cosmologies beyond-$\Lambda$CDM~\cite{Winther:2019mus,Ramachandra:2020lue,Arnold:2021xtm,Harnois-Deraps:2022bie,Brando:2022gvg,Ruan:2023mgq}. In order to train machine learning algorithms to be used for the emulation, we still need to create a large enough training, testing and validation simulation sets, however, the total number of simulations required for these are reduced by at least two orders of magnitude, where one would need around $\mathcal{O}(10^{2})$ simulations to be divided into the specified sets. While these techniques considerably reduce the number of required simulations, the problem of the computational time and complexity one MG {\it N}-body simulation takes to complete still persists.

To avoid this issue, some methods~\cite{Lombriser:2016zfz} were developed in the literature to predict the non-linear matter power spectrum in MG theories that do not need to solve for the extra scalar field equation of motion, such as the Particle-Mesh {\it N}-body code \texttt{MG-evolution}~\cite{Hassani:2020rxd}, or the \texttt{reACT} code\footnote{\hyperlink{https://github.com/nebblu/ACTio-ReACTio}{https://github.com/nebblu/ACTio-ReACTio}}~\cite{Bose:2020wch,Bose:2022vwi}. In both cases, screening mechanisms are introduced in real space by modelling the spherical collapse in a given MG theory, and then transformed to Fourier space. However, the mapping between real and Fourier space is non-trivial when screening is present, as screening depends on the density distribution in a given region, making the Fourier transformation dependent on the environment. Due to this, both codes introduce extra parameters in their screening modelling that captures the screening scale in real space and transform them to Fourier space in a phenomenological and quick way by tuning these parameters with complete\footnote{Complete in the sense that the full scalar field equation of motion is being solved consistently with the geodesic and Poisson equations.} {\it N}-body simulations. A similar approach using the spherical collapse to model screening was implemented in the COmoving Lagrangian Approximation (COLA) method~\cite{Tassev:2013pn,Tassev:2015mia,Winther:2017jof,Wright:2017dkw,Izard:2015dja,Howlett:2015hfa,Valogiannis:2016ane,Fiorini:2021dzs,Fiorini:2022srj,Wright:2022krq}, where one solves a linearized version of the equation of motion for the scalar field fluctuation in real space with two extra parameters introduced.

In view of the dependence on complete {\it N}-body simulations to model screening in non-linear predictions for the matter power spectrum, in this work we will explore a different approach, which was introduced by R. Scoccimarro in Reference~\cite{Scoccimarro:2009eu}. We will also extend the formalism introduced by Scoccimarro to a broader set of MG theories that can be described either in a covariant way, i.e., starting from a specific Lagrangian, or from a model-independent approach as the Effective Field Theory of Dark Energy (EFT of DE)~\cite{Gubitosi:2012hu,Gleyzes:2014rba}. In order to show the robustness of our results, we implemented this approximate screening approach in the publicly available \texttt{COLA-FML} code\footnote{\hyperlink{https://github.com/HAWinther/FML}{https://github.com/HAWinther/FML}}, and we validated the implementation using complete {\it N}-body simulations available in the literature. Our main results will be on the comparison of the boost function of MG theories, i.e., the ratio of the non-linear matter power spectrum from an MG theory and its GR counterpart. The boost function was recently shown~\cite{Kaushal:2021hqv,Brando:2022gvg} to be robust enough to be emulated, as it removes much of the inaccuracies present in the power spectrum from simulations and it is largely insensitive to the variation of the force resolution of the simulation, i.e., we do not need to use high resolution settings for our COLA simulations to get an accurate prediction for the boost, reducing the computational time and complexity of our simulations.

Our goal is to present a systematic procedure towards quickly generating MG non-linear predictions for the power spectrum that do not rely on the tuning of extra fitting parameters with complete {\it N}-body simulations. This will allow us to construct new emulators or extend current $\Lambda$CDM emulators to beyond-$\Lambda$CDM cosmologies, thus, fully exploring the power of emulation techniques and its use to constrain MG theories using stage-IV LSS data. The outline of this work is the following, in Section \ref{sec:method} we begin by describing the theories of gravity considered in this work in Sections~\ref{sec:method_ndgp} and \ref{sec:method_mcht}, and in Section~\ref{sec:method_scocc} we discuss the procedure introduced by Scoccimarro. In Section~\ref{sec:sims} we detail the MG {\it N}-body simulation suites used to validate our work, and in Section~\ref{sec:results} we show our results and discuss its validity with respect to its counterparts from the {\it N}-body suites and we finally conclude in Section~\ref{sec:concl}.

\section{Methodology}\label{sec:method}

In this section we will discuss the theories of gravity explored in this work, as well as summarize and discuss the screening approximation introduced by R. Scoccimarro in~\cite{Scoccimarro:2009eu}.
  
\subsection{nDGP}\label{sec:method_ndgp}

The first theory of gravity we will discuss is the so-called normal-branch Dvali-Gabadaze-Porrati (nDGP) theory~\cite{dgp}. This theory is a braneworld theory, which assumes that we live in the four-dimensional brane in a higher dimensional spacetime. Although not well motivated from first principles, nDGP has been extensively studied in the literature~\cite{fabian_ndgp1,fabian_ndgp2,Hernandez-Aguayo:2021kuh}, and serves as a test-bed for many interesting studies, due to its simplicity when introducing it to full {\it N}-body simulations. This comes from the simple form the Vainshtein mechanism~\cite{vain} is described in this theory when studying spherical collapse. The action takes the following form:
\begin{equation}\label{eq:dgp_action}
    S = \frac{1}{\kappa^{(5)}} \int \dd^{5}x \sqrt{-g^{(5)}} R^{(5)} + \int \dd^{4} \sqrt{-g} \left(\frac{1}{2\kappa}R+\mathcal{L}_{m}\right),
\end{equation}
where all quantities carrying the index $(5)$ refers to the generalization of their four-dimensional counterpart to a five-dimensional spacetime, and $\kappa = 8\pi G$, with $G$ being the Newtonian gravitational constant. Formally, when one derives the background evolution of the Universe in this theory, one encounters the fact that the action in Equation~(\ref{eq:dgp_action}) exhibits two branches to describe the expansion history: the self-accelerating branch, i.e., the branch where there is no need to introduce a Cosmological Constant to explain the late-time acceleration of the Universe, and one so-called normal branch (nDGP), in which $\Lambda$ must be introduced as well as the cross-over radius $r_{c}=G_{(5)}/2G$ needs to be tuned in order to explain the late-time acceleration. The self-accelerating branch is plagued by ghost instabilities so we only consider the normal branch in this paper. The Poisson equation in this theory, at linear order and well inside the horizon, reads:
\begin{equation}\label{eq:poi_ndgp}
    \nabla^{2}\Psi = 4 \pi \Geff a^{2}\rho_{\rm m} \delta_{\rm m},
\end{equation}
with
\begin{align}
    &\Geff = G\left(1 + \frac{1}{3\beta(a)}\right)\label{eq:Geff_ndgp},\\
    &\beta\left(a\right) = 1 - 2Hr_{c}\left(1 + \frac{\dot{H}}{3H^{2}}\right)\label{eq:beta_ndgp},
\end{align}
where a dot refers to a time derivative with respect to physical time $t$. The non-linear equations of motion are given by:
\begin{align}
    &\nabla^{2}\Psi = 4 \pi G a^{2}\rho_{\rm m} \delta_{\rm m} + \frac{1}{2}\nabla^{2}\varphi,\label{eq:poi_nl_ndgp}\\
    &\nabla^{2}\varphi + \frac{r_{c}^{2}}{3 \beta a^{2}} \left[\left(\nabla^{2}\varphi\right)^{2} - \left(\partial_{i}\partial_{j}\varphi\right)^{2}\right] = \frac{8 \pi G}{3\beta}a^{2}\rho_{\rm m } \delta_{\rm m}\label{eq:scf_nl_ndgp}.
\end{align}
One can then consider spherically symmetric solutions of Equations~(\ref{eq:poi_nl_ndgp}-\ref{eq:scf_nl_ndgp}) and find the non-linear version of Equation~(\ref{eq:Geff_ndgp})~\cite{Hernandez-Aguayo:2020kgq, fabian_ndgp1,fabian_ndgp2} as
\begin{align}\label{eq:Geff_nl_ndgp}
    &G_{\rm eff}^{\rm NL}(a,R) = G\left[1 + 2\deltaG\frac{R^{3}}{R_{V}^{3}}\left(\sqrt{1+\frac{R_{V}^{3}}{R^{3}}} - 1\right) \right],
\end{align}
where the so-called Vainshtein radius $R_V$ is given by
\begin{equation}
    \frac{R_{V}^{3}}{R^{3}} = \frac{8r_{c}^{2}H_{0}^{2}\Omega_{\rm m0} \delta_{\rm m}}{9\beta^{2}a^{3}},
\end{equation}
and
\begin{equation}
    \deltaG = \frac{\Geff - G}{G} = \frac{1}{3\beta}.
\end{equation}
Expression (\ref{eq:Geff_nl_ndgp}) is only for spherically symmetric configurations, and it is a description of how spherical collapse in real space is affected by the Vainshtein mechanism. %In the next section we will discuss another theory of gravity which we will be discussing throughout this paper.

\subsection{Horndeski Gravity}\label{sec:method_mcht}

Horndeski gravity~\cite{Horndeski:1974wa,Deffayet:2009wt,Kobayashi:2011nu} is the most general scalar-tensor theory with second-order differential equations for the metric and the scalar field. Its action is given by:
\begin{equation}
S[g_{\mu\nu},\phi]=\int\mathrm{d}^{4}x\,\sqrt{-g}\left[\sum_{i=2}^{5}\frac{1}{\kappa}{\cal L}_{i}[g_{\mu\nu},\phi]\,+\mathcal{L}_{\text{m}}[g_{\mu\nu},\psi_{\rm M}]\right]\,,\label{eq:actionhorn}
\end{equation}
where the $\mathcal{L}_{i}$ terms in the Lagrangian are:
\begin{subequations}
\begin{eqnarray}
{\cal L}_{2} & = & K(\phi,\,X)\,,\label{eq:L2}\\
{\cal L}_{3} & = & -G_{3}(\phi,\,X)\Box\phi\,,\label{eq:L3}\\
{\cal L}_{4} & = & G_{4}(\phi,\,X)R+G_{4X}(\phi,\,X)\left[\left(\Box\phi\right)^{2}-\phi_{;\mu\nu}\phi^{;\mu\nu}\right]\,,\label{eq:L4}\\
{\cal L}_{5} & = & G_{5}(\phi,\,X)G_{\mu\nu}\phi^{;\mu\nu}-\frac{1}{6}G_{5X}(\phi,\,X)\left[\left(\Box\phi\right)^{3}+2{\phi_{;\mu}}^{\nu}{\phi_{;\nu}}^{\alpha}{\phi_{;\alpha}}^{\mu}-3\phi_{;\mu\nu}\phi^{;\mu\nu}\Box\phi\right] \,. \label{eq:L5}
\end{eqnarray}
\end{subequations}
where $X\equiv -\frac{1}{2}\partial_{\mu}\phi\partial^{\mu}\phi$ is the kinetic term of the scalar field and $\psi_{\rm M}$ represents matter fields minimally coupled to gravity. Equation~(\ref{eq:actionhorn}) follows the same notations and conventions with respect to the Horndeski functions $G_{i}$ as~\cite{Bellini:2014fua} and \hiclass\footnote{\hyperlink{https://github.com/miguelzuma/hi_class_public}{https://github.com/miguelzuma/hi\_class\_public}}~\cite{Zumalacarregui:2016pph,Bellini:2019syt} a public Einstein-Boltzmann code, which will be extensively used in this paper.

The background equations of motion for the action in Equation~(\ref{eq:actionhorn}) reads:
\begin{subequations}
\begin{align}
    H^{2} &= \frac{\kappa}{3} \left( \sum_{\alpha}\rho_{\alpha} + \rho_{\mathrm{DE}} \right), \label{eq:H2}\\
    \dot{H} &= - \frac{\kappa}{2} \left[ \sum_{\alpha}\left(\rho_{\alpha} + p_{\alpha}\right) + \rho_{\mathrm{DE}} + p_{\mathrm{DE}}\right],\label{eq:Hprime}
\end{align}
\end{subequations}
where a dot represents a derivative with respect to the physical time $t$, $\alpha$ runs over all matter species\footnote{With matter we also mean relativistic species such as photons and massless neutrinos.}, and
\begin{subequations}
\begin{align}
\kappa \mathcal{\rho_{\text{DE}}} &\equiv  -K+2X\left(K_{X}-G_{3\phi}\right)+6\dot{\phi}H\left(XG_{3X}-G_{4\phi}-2XG_{4\phi X}\right)\label{eq:rhoDE}\\
 &+12H^{2}X\left(G_{4X}+2XG_{4XX}-G_{5\phi}-XG_{5\phi X}\right)+4\dot{\phi}H^{3}X\left(G_{5X}+XG_{5XX}\right)\, \nonumber,\\
\kappa p_{\text{DE}}&\equiv K-2X\left(G_{3\phi}-2G_{4\phi\phi}\right)+4\dot{\phi}H\left(G_{4\phi}-2XG_{4\phi X}+XG_{5\phi\phi}\right)\label{eq:pDE}\\
 & -M_{*}^{2}\alpha_{\text{B}}H\frac{\ddot{\phi}}{\dot{\phi}}-4H^{2}X^{2}G_{5\phi X}+2\dot{\phi}H^{3}XG_{5X}\, . \nonumber 
\end{align}
\end{subequations}
The dark energy background quantities satisfy the usual conservation equation:
\begin{subequations}
\begin{align}
    \dot{\rho}_{\mathrm{DE}} &= -3aH\left(\rho_{\mathrm{DE}}+p_{\mathrm{DE}}\right)\label{eq:rhoprimeDE},\\
    w_{\mathrm{DE}} &\equiv \frac{p_{\mathrm{DE}}}{\rho_{\mathrm{DE}}}\label{eq:wDE}.
\end{align}
\end{subequations}
One can see that in Equation~(\ref{eq:pDE}), there is a non-trivial quantity, $\bra$, in its definition. This is the so-called braiding function first introduced in~\cite{Bellini:2014fua} alongside other 3 time-dependent functions which are defined as:
\begin{subequations}
\begin{align}
M_{*}^{2}\equiv & 2\left(G_{4}-2XG_{4X}+XG_{5\phi}-\dot{\phi}HXG_{5X}\right),\label{eq:planckmass}\\
HM_{*}^{2}\alpha_{\textrm{M}}\equiv & \frac{\mathrm{d}}{\mathrm{d}t}M_{*}^{2},\label{eq:run_smg}\\
H^{2}M_{*}^{2}\alpha_{\textrm{K}}\equiv & 2X\left(K_{X}+2XK_{XX}-2G_{3\phi}-2XG_{3\phi X}\right)+\label{eq:kin_smg}\\
 & +12\dot{\phi}XH\left(G_{3X}+XG_{3XX}-3G_{4\phi X}-2XG_{4\phi XX}\right)+\nonumber \\
 & +12XH^{2}\left(G_{4X}+8XG_{4XX}+4X^{2}G_{4XXX}\right)-\nonumber \\
 & -12XH^{2}\left(G_{5\phi}+5XG_{5\phi X}+2X^{2}G_{5\phi XX}\right)+\nonumber \\
 & +4\dot{\phi}XH^{3}\left(3G_{5X}+7XG_{5XX}+2X^{2}G_{5XXX}\right),\nonumber \\
HM_{*}^{2}\alpha_{\textrm{B}}\equiv & 2\dot{\phi}\left(XG_{3X}-G_{4\phi}-2XG_{4\phi X}\right)+\label{eq:bra_smg}\\
 & +8XH\left(G_{4X}+2XG_{4XX}-G_{5\phi}-XG_{5\phi X}\right)+\nonumber \\
 & +2\dot{\phi}XH^{2}\left(3G_{5X}+2XG_{5XX}\right),\nonumber \\
M_{*}^{2}\alpha_{\textrm{T}}\equiv & 2X\left(2G_{4X}-2G_{5\phi}-\left(\ddot{\phi}-\dot{\phi}H\right)G_{5X}\right).\label{eq:ten_smg}
\end{align}
\end{subequations}

These are dimensionless quantities that capture all the features introduced by the dark energy scalar field at a \emph{linear} level, and we refer the reader to Section~3.1 of~\cite{Bellini:2014fua} for the physical meaning of each quantity. Under the quasi-static approximation~\cite{Pace:2020qpj,Bellini:2014fua,Brando:2021jga} and at linear order, we can write the Poisson equation for general Horndeski theories as:
\begin{equation}
    \nabla^{2}\Psi = 4 \pi G_{\rm eff} a^{2} \rho_{\rm m}\delta_{\rm m},
\end{equation}
with
\begin{equation}\label{eq:Geff_horn}
    G_\mathrm{eff} = G \left(1 + \frac{ \csN \left( 2- 2\mpl + 2\ten \right) + \left( \bra + 2 \run -2\ten + \bra\ten \right)^{2}}{2\csN \mpl}\right),
\end{equation}
where $\csN$ is the numerator of the scalar field speed of sound squared defined in~\cite{Zumalacarregui:2016pph}.
However, in the present work, we will not cover general Horndeski theories, as we are interested in the minimal case in which the Vainshtein mechanism is present. That is, we will be looking at the subset:
\begin{equation}\label{eq:MCHT}
    \bra \neq 0, \ \ \ \run = \ten = 0, \ \ \ \forall \kin.
\end{equation}
This minimal set can be translated in terms of the Horndeski functions $G_{i}$ as:
\begin{align}
    &G_{4} = \frac{1}{2}R, \ \ \ G_{5}=0,\nonumber \\
    &K = K\left(\phi, X\right), \ \ \ G_{3} = G_{3}\left(\phi, X\right).
\end{align}
This is a subset of Horndeski theories that are \emph{minimally coupled to the metric tensor} (MCHT) which satisfy the bound set by the merger of binary neutron stars on the speed of gravitational waves~\cite{LIGOScientific:2017vwq}, i.e., $c_{\rm GW}=c$, and, Equation~(\ref{eq:Geff_horn}) in these theories becomes:
\begin{equation}\label{eq:Geff_mcht_hc}
    \Geff = G\left(1 + \frac{\bra^{2}}{2\csN}\right).
\end{equation}

In the next section we will discuss the non-linear equations of motion for these theories, writing all equations of motion in terms of the property function $\bra$ or functions defined from it.

\subsubsection{Spherical Collapse}\label{sec:method_sph_col}

In order to understand how MCHT contributes to the non-linear behavior of matter, we will first write the equations of motion for the metric perturbations, as well as for the \emph{dimensionless} scalar field fluctuation defined as:
\begin{equation}\nonumber
    V_{X} = \frac{\delta \phi}{\dot{\phi}}.
\end{equation}
We begin with the perturbed FLRW line element in the Newtonian gauge:
\begin{equation}\label{eq:met_pert_flrw}
    \dd s^{2} = - \left( 1 + 2\Phi \right)\dd t^{2} + a^{2}(t)\left( 1 - 2 \Psi \right)\delta_{ij}\dd x^{i} \dd x^{j},
\end{equation}
which takes the same form as the line element in~\cite{Cusin:2017mzw}. Using Equations~(\ref{eq:MCHT}), (\ref{eq:met_pert_flrw}) and Equation~(3.23) of~\cite{Cusin:2017mzw}, we arrive at:
\begin{subequations}
\begin{align}%\label{eq:eom_mcht}
    &\bra H \partial^{2} \Psi = 2 C_{2}H^{2}\partial^{2} V_{X} + \frac{\bra H}{a^{2}} V_{X}^{(2)}\label{eq:Vx_mcht},\\
    &\partial^{2}\Psi = \frac{\kappa}{2}a^{2}\delta \rho_{\rm m} - \frac{1}{2}\bra H \partial^{2}V_{X}\label{eq:poi_mcht},
\end{align}
\end{subequations}
with the following definitions:
\begin{subequations}
\begin{align}
    &C_{2} \equiv - \frac{\bra}{2} - \frac{1}{2H^{2}}\frac{\dd}{\dd t}\left(\bra H\right) - \frac{1}{H^{2}}\left( \rho_{\rm DE} + p_{\rm DE}\right),\\
    &V_{X}^{(2)} \equiv \left(\partial^{2}V_{X}\right)^{2} - \left(\partial_{i}\partial_{j}V_{X}\right)^{2},\\
    &\rho_{\rm DE} + p_{\rm DE} = 2X\left( G_{2X} - 2 G_{3\phi} + 3H\dot{\phi} \right) - \bra H \frac{\ddot{\phi}}{\dot{\phi}}.
\end{align}
\end{subequations}
We can also find the expression for the linear effective gravitational coupling, Equation~(\ref{eq:Geff_mcht_hc}), from Equations~(\ref{eq:Vx_mcht}-\ref{eq:poi_mcht}):
\begin{equation}\label{eq:Geff_nl_mcht}
    \Geff = G\left(1 - \frac{\bra^{2}}{C_{2} + \bra^{2}}\right).
\end{equation}
Equations~(\ref{eq:Vx_mcht}-\ref{eq:poi_mcht}) are general equations for cubic Horndeski theories minimally coupled to the metric tensor that incorporate the Vainshtein screening. This is due to the fact that in Equation~(\ref{eq:Vx_mcht}) we have the presence of the non-linear term $V_{X}^{(2)}$, which effectively decouples the scalar degree of freedom from the matter density fluctuations on small scales in dense environments, hence, shielding the fifth force introduced by the extra degree of freedom. Note, however, that contrary to other screening mechanisms, in theories with Vainshtein screening the scalar field is still dynamical, only the fifth force is suppressed. To better understand how screening occurs in these theories, we will now consider the case where density perturbations are spherically symmetrically distributed. We will first combine Equations~(\ref{eq:Vx_mcht}-\ref{eq:poi_mcht}) in order to find a governing equation of motion for the scalar field fluctuation that only depends on the matter density perturbations:
\begin{align}\label{eq:dphi_vainsh}
    \partial^{2}V_{X} + \frac{2\bra}{H\left( 4C_{2} + \bra^{2} \right)} \frac{V_{X}^{(2)}}{a^{2}} = \frac{2\bra}{H\left( 4C_{2} + \bra^{2} \kappa \right)} \frac{\kappa}{2} a^{2} \delta \rho_{\rm m}.
\end{align}
In order to make this equation more compact, and to make it similar to its counterpart in nDGP, Equation~(\ref{eq:scf_nl_ndgp}), we will introduce the following function:
\begin{equation}
    \frac{1}{3\beta_{1}} \equiv \frac{2\bra H}{\left( 4C_{2} + \bra^{2} \right)},
\end{equation}
which then allows us to rewrite Equation~(\ref{eq:dphi_vainsh}) as:
\begin{align}
    \partial^{2} V_{X} + \frac{1}{3\beta_{1}a^{2}}V_{X}^{(2)} = \frac{1}{6\beta_{1}}\kappa a^{2} \delta \rho_{\rm m},
\end{align}
which in radial coordinates becomes:
\begin{equation}
    \frac{1}{r^{2}}\frac{\dd}{\dd r}\left(r^{2}\frac{\dd V_{X}}{\dd r}\right) + \frac{2}{3\beta_{1}a^{2}} \frac{1}{r^{2}}\frac{\dd }{\dd r}\left[r \left(\frac{\dd V_{X}}{\dd r}\right)^{2}\right] = \frac{\kappa}{6\beta_{1}}a^{2}\rho_{\rm m}\delta_{\rm m}.
\end{equation}
This equation can be integrated once, leading to:
\begin{equation}\label{eq:V_X_r_vain}
    \frac{\dd V_{X}}{\dd r} + \frac{2}{3\beta_{1}a^{2}} \frac{1}{r}\left(\frac{\dd V_{X}}{\dd r}\right)^{2} = \frac{1}{3\beta_{1}}a^{2}\frac{G M(r)}{r^{2}},
\end{equation}
where $M(r) = 4\pi \int_{0}^{r}\sigma^{2}\rho_{\rm m}\delta_{\rm m}(\sigma)\dd \sigma$ is the enclosed mass within the radius $r$. It is easy to see that Equation~(\ref{eq:V_X_r_vain}) is a second order algebraic equation for the variable $\dd V_{X}/\dd r$, and assuming a top-hat density distribution, % for the profile of the enclosed mass within $r=R$
we find the following expression at $r=R$: 
\begin{equation}\label{eq:dvx_dr}
    \frac{\dd V_{X}}{\dd r} = \frac{4a^{2}}{3 \beta_{1}} \frac{R^{3}}{2R_{V}^{3}}\left[ \sqrt{1+\frac{R_{V}^{3}}{R^{3}}} -1 \right]\frac{G M}{R^{2}}.
\end{equation}
%which is valid outside the top-hat density distribution, and the physically relevant solution to us. 
Differentiating Equation~(\ref{eq:dvx_dr}) with respect to $r$ we arrive at:
\begin{equation}\label{eq:d2Vx_dr2}
    \frac{\dd^{2}V_{X}}{\dd r^{2}} = \frac{\kappa}{2} \frac{4a^{2}}{3\beta_{1}}\frac{R^{3}}{2R_{V}^{3}}\left[\sqrt{1+\frac{R^{3}_{V}}{R^{3}}} - 1\right]\rho_{\rm m }\delta_{\rm m},
\end{equation}
and we can use this result in Equation~(\ref{eq:poi_mcht}) to write:
\begin{equation}\label{eq:poi_vain_of_r}
    \partial^{2}\Psi = \frac{\kappa}{2} a^{2} \rho_{\rm m}\delta_{\rm m} \left[ 1 + 2 \frac{\delta G}{G} \frac{R^{3}}{R_{V}^{3}} \left( \sqrt{1+\frac{R_{V}^{3}}{R^{3}}}-1\right) \right],
\end{equation}
with
\begin{equation}
    \frac{\delta G}{G} = - \frac{\bra H }{6 \beta_{1}}.
\end{equation}
Similarly to Equation~(\ref{eq:Geff_nl_ndgp}) for nDGP, we can find an expression for the non-linear version of Equation~(\ref{eq:Geff_nl_mcht}):
\begin{equation}\label{eq:poi_sph_real}
    \partial^{2}\Psi = \frac{\kappa}{2}G_{\rm eff}^{\rm NL}(a,R)a^{2}\rho_{\rm m} \delta_{\rm m}
\end{equation}
where
\begin{equation}\label{eq:Geff_nl}
    G_{\rm eff}^{\rm NL} = G \left[ 1 + 2 \frac{\delta G}{G} \frac{R^{3}}{R_{V}^{3}} \left( \sqrt{1+\frac{R_{V}^{3}}{R^{3}}}-1\right) \right],
\end{equation}
and
\begin{equation}
    \frac{R_{V}^{3}}{R^{3}} = \frac{1}{2}~\frac{8H_{0}^{2}\Omega_{\rm m0} \delta_{\rm m}}{9\beta_{1}^{2}a^{3}}.
\end{equation}
Equation~(\ref{eq:Geff_nl}) allows us to understand the two regimes in which the fifth force introduced by the scalar field acts. When the matter density perturbations are large, we have that $R/R_{V}\ll 1$, and, in this case, $G_{\rm eff}^{\rm NL} \to G$, while for small linear perturbations $G_{\rm eff}^{\rm NL} \to G_{\rm eff}$. Equation~(\ref{eq:Geff_nl}) is still a simplification to the question of how modified gravity acts at small scales, as we have derived this result using a spherically symmetric profile for the scalar field fluctuations in real space. This result has been found and discussed extensively in the literature~\cite{Cusin:2017mzw,Lombriser:2016zfz,Hassani:2020rxd,Hernandez-Aguayo:2020kgq,kk_silva}, as well as it has been used to approximately implement screening effects in {\it N}-body simulations~\cite{Hassani:2020rxd}. In the next section, however, we will discuss an alternate procedure to introduce screening in {\it N}-body simulations.

\subsection{Scoccimarro's Prescription}\label{sec:method_scocc}

In this section we will review the method introduced by R. Scoccimarro in~\cite{Scoccimarro:2009eu} to model screening in nDGP gravity, and extend it to Horndeski theories characterized by (\ref{eq:MCHT}). This idea is based on a way to compute an effective gravitational constant in modified gravity theories that exhibit Vainshtein screening, which will depend on time \emph{and} scale, i.e., $\Geff\left(z,k\right)$. We have seen in the previous section how this function is computed in real space by studying the spherical collapse, i.e. Equations~(\ref{eq:Geff_nl_ndgp}) and (\ref{eq:Geff_nl}). We will present our results in a unifying description aiming to aid the efforts to introduce modified gravity effects in the non-linear matter power spectrum with the precision and speed required for stage-IV LSS surveys.

\begin{figure}[h] 
\centering
\includegraphics[width=1.\textwidth]{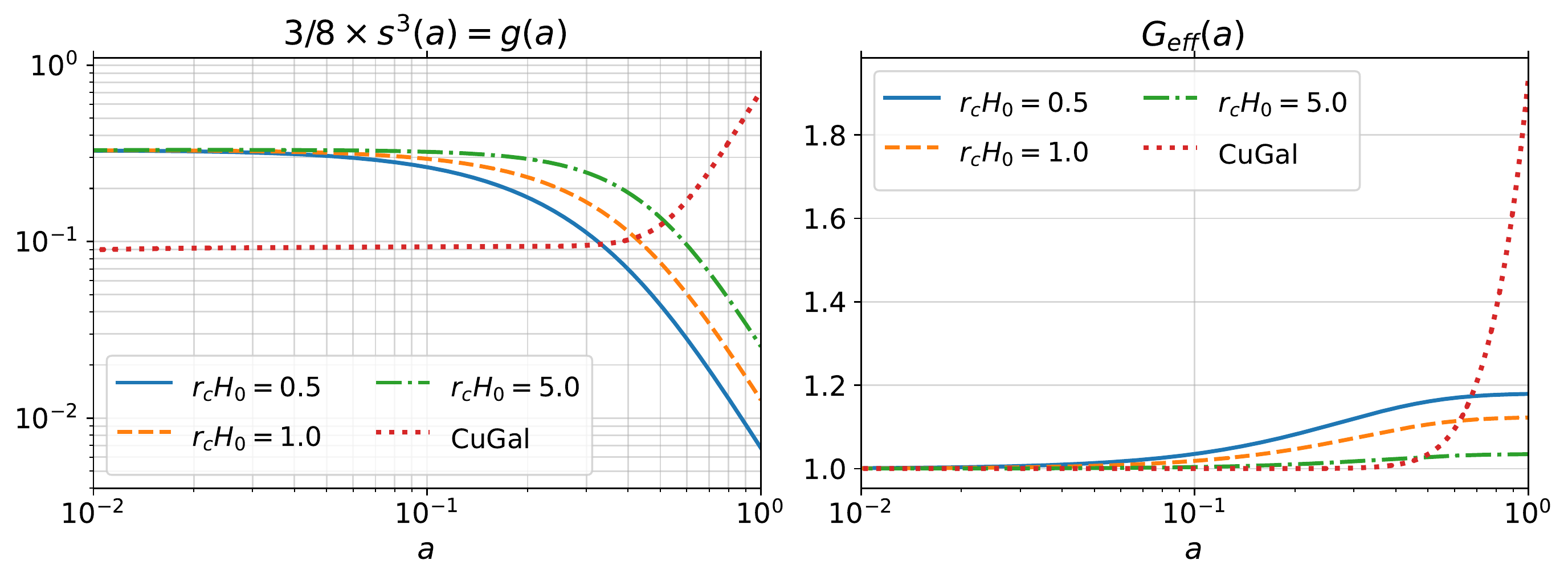}
\caption{\textbf{Left:} Evolution of the non-linearity function $g(a)$ as a function of scale factor. \textbf{Right:} Evolution of the linear and time-dependent only effective gravitational coupling $G_{\rm eff}$, Equation~(\ref{eq:Geff_horn}) also as a function of the scale factor. Solid blue curves are for nDGP gravity with $r_{c}H_{0}=0.5$, dashed orange curves $r_{c}H_{0}=1.0$, dotted-dashed green curves $r_{c}H_{0}=5.0$ and dotted red curves for the Cubic Galileon (CuGal) model.}
\label{fig:ndgp_bg}
\end{figure}

We begin by combining Equations~(\ref{eq:Vx_mcht}-\ref{eq:poi_mcht}) into one equation as:
\begin{equation}\label{eq:rs_bra}
    \partial^{2}\Psi - \frac{\kappa}{2}a^{2}\rho_{\rm m}\Geff \delta_{\rm m} = -\deltaG V_{X}^{(2)}.
\end{equation}
In Fourier space, Equation~(\ref{eq:rs_bra}) can be written as:
\begin{equation}
    -\left(k^{2}\Psi + \frac{\kappa}{2}a^{2}\rho_{\rm m}\Geff\delta_{\rm m}\right) = - \deltaG \int \frac{\dd^{3}k_{1}\dd^{3}k_{2}}{\left(2\pi\right)^{3}}\delta_{\rm D}\left(\vec{k}_{12}-\vec{k}\right)\left[k_{1}^{2}k_{2}^{2}-\left(\vec{k}_{1}\cdot \vec{k}_{2}\right)^{2}\right]V_{X}(\vec{k}_{1})V_{X}(\vec{k}_{2})
\end{equation}
where $\vec{k}_{12}=\vec{k}_{1}+\vec{k}_{2}$. Using Equation~(\ref{eq:poi_mcht}) we can write in Fourier space:
\begin{equation}
    V_{X} = -\frac{1}{k^{2}} \frac{2}{\bra H} \left[ k^{2}\Psi + \frac{\kappa}{2}a^{2}\rho_{\rm m}\delta_{\rm m} \right],
\end{equation}
which brings us to:
\begin{align}\label{eq:rs_bra_fourier_full}
    -\left[k^{2}\Psi + \frac{\kappa}{2}a^{2}\rho_{\rm m} \delta_{\rm m}\right] = \deltaG\Bigg\lbrace \frac{\kappa}{2}\rho_{\rm m}\delta_{\rm m} &+ \frac{4}{\bra^{2}H^{2}} \int \frac{\dd^{3}k_{1}\dd^{3}k_{2}}{\left(2\pi\right)^{3}}\delta_{\rm D}\left(\vec{k}_{12}-\vec{k}\right)\left[1-\left(\vec{\hat{k}}_{1}\cdot \vec{\hat{k}}_{2}\right)^{2}\right] \nonumber \\ 
    &\times \left[k_{1}^{2}\Psi + \frac{\kappa}{2}a^{2}\rho_{\rm m}\delta_{\rm m}\right] \left[k_{2}^{2}\Psi + \frac{\kappa}{2}a^{2}\rho_{\rm m}\delta_{\rm m}\right] \Bigg\rbrace.
\end{align}
Equation~(\ref{eq:rs_bra_fourier_full}) represents the full solution of the Poisson potential in the quasi-static approximation. However, performing the integral in Equation~(\ref{eq:rs_bra_fourier_full}) is highly non-trivial, and a simplification of it is to consider the spherically symmetric case for the non-local kernel $[1-(\vec{\hat{k}}_{1}\cdot \vec{\hat{k}}_{2})^{2}]= 2/3$, which when Fourier transformed back to real space restores to the result in Equation~(\ref{eq:poi_sph_real}), which we will now write as:
\begin{equation}\label{eq:poi_w_s3}
    \partial^{2}\Psi = \frac{\kappa}{2}a^{2}\rho_{\rm m} \left[ \delta_{\rm m} + 2 \deltaG \frac{1}{s^{3}}\left(\sqrt{1+s^{3}\delta_{\rm m}}-1\right)\right],
\end{equation}
where we have introduced
\begin{equation}\label{eq:s_3_bra}
    s^{3} = \frac{1}{2}~\frac{8H_{0}^{2}\Omega_{\rm m0}}{9\beta_{1}^{2}a^{3}}.
\end{equation}
This quantity is related to the non-linear scale where the Vainshtein mechanism kicks in, as one can compare Equation~(\ref{eq:s_3_bra}) with Equation~(\ref{eq:Geff_nl}) and see that $s^{3}$ is related to $R^{3}/R_{V}^{3}$. Equation~(\ref{eq:s_3_bra}) is valid for Horndeski theories that satisfy the condition in Equation~(\ref{eq:MCHT}), and it can be compared with the expression for nDGP theories:
\begin{equation}\label{eq:s_3_ndgp}
    s^{3} = \frac{8r_{c}^{2}H_{0}^{2}\Omega_{\rm m0}}{9\beta^{2}a^{3}}.
\end{equation}
 In Figure~\ref{fig:ndgp_bg} we show the time evolution of the quantity $g(a)=3/8 s^{3}(a)$ for nDGP gravity and Cubic Galileon on the left hand side, while on the right hand side we plot the effective gravitational constant $G_{\rm eff}(a)$, Equations~(\ref{eq:Geff_horn}) and (\ref{eq:Geff_ndgp}), in the cosmological background.
 %%%%%%%%%%%%%%%%
%%%%%%%%%%%%%%%%
In the left plot of Figure~\ref{fig:ndgp_bg}, we can see that all nDGP curves at early times converge to their asymptotic value, $g(a)\sim 1/3$, while at late times they will differ between each other as each has a different value of $r_{c}H_{0}$. With respect to the orange dashed curve, corresponding to $r_{c}H_{0}=1.0$, our plot is in agreement with the pink dashed curve found in the plot of Figure 3 in~\cite{fabian_ndgp2}\footnote{Note that the plotted quantity in that paper is the inverse of the quantity plotted in this work. However, one can simply invert the orange dashed curve late time behavior to find that their asymptotic value is in agreement. Also, the nDGP gravity parameter considered for the pink dashed curve in that paper is $r_{c}H_{0}\approx 0.72$, which is close enough to the value of the dashed orange curve in Figure~\ref{fig:ndgp_bg} to get an estimate on the order of magnitude of the non-linear function at late times.}. The behavior for the Cubic Galileon, however, is opposite at late-times than its nDGP counterpart. 

Equation~(\ref{eq:rs_bra_fourier_full}) correctly captures the non-local and non-linear nature of the connection between the Poisson potential and the density perturbation. Therefore, Scoccimarro introduces a technique to regain some non-local aspects of the true solution by employing a ressumation technique on the two-point propagator in Fourier space, which will capture non-local corrections beyond the spherical symmetric approximation. The starting point is to write the response of the Poisson potential in Fourier space:
\begin{equation}\label{eq:poi_pot_resp}
    \Psi\left(\vec{k}\right) = \left\langle \frac{\mathcal{D}\Psi\left(\vec{k}\right)}{\mathcal{D}\delta_{\rm m}\left(\vec{k}^{\prime}\right)} \right\rangle \delta_{\rm m}\left( \vec{k} \right) + \frac{1}{2!}\left\langle \frac{\mathcal{D}^{2}\Psi\left(\vec{k}\right)}{\mathcal{D}\delta_{\rm m}\left(\vec{k}_{1}\right)\mathcal{D}\delta_{\rm m}\left(\vec{k}_{2}\right)} \right\rangle \delta_{\rm m}\left(\vec{k}_{1}\right)\delta_{\rm m}\left(\vec{k}_{2}\right) + \dots ,
\end{equation}
where $\mathcal{D}$ is a functional derivative, and $\langle \dots \rangle$ represent statistical averages. The two-point propagator of the modified Poisson equation, in the context of renormalized perturbation theory~\cite{Crocce:2005xy}, is then the linear coefficient of the previous expansion:
\begin{equation}\label{eq:lin_coeff}
    \Gamma_{\Psi}\left(\vec{k}\right) \delta_{\rm D}\left(\vec{k}-\vec{k}^{\prime}\right)= \left\langle \frac{\mathcal{D}\Psi\left(\vec{k}\right)}{\mathcal{D}\delta_{\rm m}\left(\vec{k}^{\prime}\right)} \right\rangle.
\end{equation}
From Equation~(\ref{eq:poi_w_s3}) we find the asymptotic behavior of $\Gamma_{\Psi}\left(\vec{k}\right)$:
\begin{subequations}\label{eq:Gamma_Psi_asymp}
\begin{align}
    &\Gamma_{\Psi}\left(k R_{V} \ll 1 \right) \simeq -\frac{1}{k^{2}}\frac{\kappa}{2}a^{2}\rho_{\rm m}\Geff, \\
    &\Gamma_{\Psi}\left(k R_{V} \gg 1 \right) \simeq -\frac{1}{k^{2}}\frac{\kappa}{2}a^{2}\rho_{\rm m} .
\end{align}
\end{subequations}
Equation~(\ref{eq:poi_pot_resp}) is an expansion for statistical averages such as the power spectrum, therefore, we can naively replace $\delta_{\rm m}\to \Delta$ in Equation~(\ref{eq:poi_w_s3}), which allows us to estimate the linear coefficient:
\begin{equation}\label{eq:Gamma_Psi_Delta}
    \Gamma_{\Psi}\left(k,z\right) \approx - \frac{1}{k^{2}}\frac{\kappa}{2} a^{2}\rho_{\rm m} \left[ \frac{\delta G / G}{\sqrt{1+s^{3}\Delta\left(k,z\right)}} \right],
\end{equation}
where $\Delta(k)$ is the square root of the dimensionless power spectrum:
\begin{equation}
    \Delta^2\left(k,z\right) = \frac{k^{3}P_{\rm m }\left(k,z\right)}{2\pi^{2}}.
\end{equation}
As argued in~\cite{Scoccimarro:2009eu}, in order to get an effective resummation scheme it is reasonable to use the linear order version of Equation~(\ref{eq:rs_bra_fourier_full}) and introduce a new function, $M(k)$, to capture non-local features that are averaged out:
\begin{equation}\label{eq:poi_w_M}
    \left[k^{2}\Psi\left(k,z\right) + \frac{\kappa}{2}a^{2}\rho_{\rm m} \delta_{\rm m}\left(k,z\right)\right] = -\deltaG\frac{\kappa}{2}a^{2}\rho_{\rm m}\delta\left(k,z\right)M\left(k,z\right),
\end{equation}
which then leads us to the following expression:
\begin{equation}\label{eq:Gamma_Psi_M}
    \Gamma_{\Psi}\left(k,z\right) = - \frac{1}{k^{2}}\frac{\kappa}{2} a^{2}\rho_{\rm m} \left[\frac{\delta G \left(z\right)}{G}  \times M\left(k,z\right) \right].
\end{equation}
The function $M(z,k)$ should describe non-local and non-linear features of the Vainshtein screening on \emph{average}, and the $M(k,z)$ function must have the following asymptotic behavior:
\begin{equation}\label{eq:M_asymp}
    M\left(k \ll 1\right) \to 1, \ \ \ M\left(k \gg 1\right) \to 0,
\end{equation}
otherwise Equation~(\ref{eq:Gamma_Psi_asymp}) is not recovered. Inspection of Equation~(\ref{eq:poi_w_M}) then shows us that the result of this procedure will give an expression for $\Geff$ that depends on time and scale:
\begin{equation}\label{eq:Geff_of_k_z}
    \Geff\left(z,k\right) = G\left( 1 + \frac{\delta G \left(z\right)}{G} M\left(z, k\right)\right).
\end{equation}
In order to correctly model the Vainshtein mechanism, we need to go beyond the initial linear order version of Equation~(\ref{eq:rs_bra_fourier_full}), and for this we rewrite Equation~(\ref{eq:rs_bra_fourier_full}) using Equation~(\ref{eq:Gamma_Psi_M}):
\begin{align}\label{eq:M_full}
    \left(M-1\right)\delta\left(\vec{k}\right) = - \frac{3}{8}s^{3} \int &\frac{\dd^{3}k_{1}\dd^{3}k_{2}}{\left(2\pi\right)^{3}}\delta_{\rm D}\left(\vec{k}_{12}-\vec{k}\right)\left[1-\left(\vec{\hat{k}}_{1}\cdot \vec{\hat{k}}_{2}\right)^{2}\right]  \nonumber \\ 
    &\times \delta_{\rm m}\left(\vec{k}_{1}\right)M\left(\vec{k}_{1}\right)\delta_{\rm m}\left(\vec{k}_{2}\right)M\left(\vec{k}_{2}\right).
\end{align}
If we multiply Equation~(\ref{eq:M_full}) by $\delta\left(\vec{k}^{\prime}\right)$, then take the statistical average and Fourier transform it back to real space we are left with:
\begin{align}\label{eq:xi_min_chi}
    \xi\left(r\right)-\chi\left(r\right) = - \frac{3}{8}s^{3} \int & \frac{\dd^{3}k_{1}\dd^{3}k_{2}}{\left(2\pi\right)^{3}}\left[1-\left(\vec{\hat{k}}_{1}\cdot \vec{\hat{k}}_{2}\right)^{2}\right] B\left(\vec{k}_{1}, \vec{k}_{2}, \vec{k}_{3}\right) \nonumber\\
    & \times M\left(\vec{k}_{1}\right)M\left(\vec{k}_{2}\right)e^{-i \vec{k}_{12}\cdot \vec{r}},
\end{align}
where $B$ is the matter bispectrum, i.e. the Fourier transform of the three-point correlation function, and $\xi, \chi$ are the correlation functions:
\begin{equation}\label{eq:chi_xi}
    \xi\left(r\right) = \int \frac{\dd^{3}k}{\left(2\pi\right)^{3}}P\left(k\right)e^{-i \vec{k}\cdot\vec{r}}, \ \ \ \chi\left(r\right) = \int \frac{\dd^{3}k}{\left(2\pi\right)^{3}}P\left(k\right)M\left(k\right)e^{-i \vec{k}\cdot\vec{r}}.
\end{equation}
Now, from Equation~(\ref{eq:xi_min_chi}) we can see that in order to compute non-Gaussian corrections induced by modified gravity on small scales we need a prescription for the bispectrum. As the Vainshtein scale lies in the non-linear regime, we need a formula for the bispectrum on small scales, and one can use a fitting formula for the standard gravity bispectrum, such as the one considered in~\cite{Scoccimarro:1997st} or its updated formula developed in~\cite{Gil-Marin:2011jtv}. In both cases the fitting formula starts from a formula analogous to the tree-level expression for the bispectrum:
\begin{align}\label{eq:B_fit}
    B_{\rm fit}\left(\vec{k}_{1}, \vec{k}_{2}, \vec{k}_{3}, z\right) = 2 F_{2}\left(\vec{k}_{1}, \vec{k}_{2}, z\right) &P\left(\vec{k}_{1}, z\right)P\left(\vec{k}_{2}, z\right) + 2 F_{2}\left(\vec{k}_{2}, \vec{k}_{3}, z\right) P\left(\vec{k}_{2}, z\right)P\left(\vec{k}_{3}, z\right) \nonumber\\
    &+ 2 F_{2}\left(\vec{k}_{3}, \vec{k}_{1}, z\right) P\left(\vec{k}_{3}, z\right)P\left(\vec{k}_{1}, z\right),
\end{align}
however, here, the $F_{2}$ kernel is given by:
\begin{align}\label{eq:F2_kernel}
    F_{2}\left(\vec{k}_{1}, \vec{k}_{2}, z\right) = \frac{5}{7}a\left(k_{1}, z\right)a\left(k_{2}, z\right)& + \frac{1}{2}\left(\vec{\hat{k}}_{1}\cdot\vec{\hat{k}}_{2}\right)\left(\frac{k_{1}}{k_{2}}+\frac{k_{2}}{k_{1}}\right)b\left(k_{1}, z\right)b\left(k_{2}, z\right)\nonumber\\
    &+ \frac{2}{7}\left(\vec{\hat{k}}_{1}\cdot\vec{\hat{k}}_{2}\right)^{2}c\left(k_{1}, z\right)c\left(k_{2}, z\right),
\end{align}
and the matter power spectrum $P\left(k\right)$ is the \emph{non-linear} power spectrum, which can be computed using parametrized prescriptions such as \texttt{halofit}~\cite{Bird:2011rb,Takahashi:2012em} and \texttt{HMCode}~\cite{Mead:2020vgs} or emulators, such as Euclid Emulator 2~\cite{Euclid:2020rfv}\footnote{\hyperlink{https://github.com/miknab/EuclidEmulator2}{https://github.com/miknab/EuclidEmulator2}} and Bacco~\cite{Angulo:2020vky}\footnote{\hyperlink{https://baccoemu.readthedocs.io/en/latest/}{https://baccoemu.readthedocs.io/en/latest/}}. The time and scale dependent functions appearing in Equation~(\ref{eq:F2_kernel}) are given by:
\begin{subequations}\label{eq:abc_funct_fit}
\begin{align}
    &a\left(n, k\right) = \frac{1 + \sigma_{8}^{a_{6}}(z) \left[ 0.7Q_{3}\left(n\right) \right]^{1/2}\left(qa_{1}\right)^{n+a_{2}}}{1+\left(qa_{1}\right)^{n+a_{2}}},\label{eq:a_fit}\\
    &b\left(n, k\right) = \frac{1 + 0.2a_{3}\left(n+3\right)\left(qa_{7}\right)^{n+3+a_{8}}}{1+\left(qa_{7}\right)^{n+3+a_{8}}},\label{eq:b_fit}\\
    &c\left(n, k\right) = \frac{1+4.5a_{4}/[1.5+\left(n+3\right)^{4}]\left(qa_{5}\right)^{n+3+a_{9}}}{1+\left(qa_{5}\right)^{n+3+a_{9}}}\label{eq:c_fit},
\end{align}
\end{subequations}
where
\begin{align}
    &Q_{3}\left(n\right) = \frac{4-2^{n}}{1+2^{n+1}}, \ \ \ n = \frac{\dd \ln P_{\rm lin}\left(k\right)}{\dd \ln k},
\end{align}
and $q = k/k_{\rm NL}$, with $k_{\rm NL}$ being the non-linear wave-number where the dimensionless linear matter power spectrum is unity:
\begin{equation}
    \frac{k_{\rm NL}^{3}P_{\rm lin}\left(k_{\rm NL}\right)}{2 \pi^{2}} = 1.
\end{equation}
In the original prescription used in~\cite{Scoccimarro:1997st} the values for the $a_{i}$ parameters are summarized in Table~\ref{tab:scocc_fit}. It is worth noting that the fit was performed without a validation to a maximum redshift, while Reference~\cite{Gil-Marin:2011jtv} performed the same fit using higher resolution simulations and re-calibrated the values for the parameters found in Table~\ref{tab:hgm_fit}, which are valid up to redshift $z=1.5$. Therefore, for the results shown in the next sections we used the parameter values presented in Table~\ref{tab:hgm_fit}.
\begin{table}[h]
    \begin{center} 
        \begin{tabular}{l|l|l} 
            \toprule
            \midrule
            $a_{1}=0.25$ & $a_{2}=3.5$ & $a_{3}=2$ \\
            $a_{4}=1$ & $a_{5}=2$ & $a_{6}=-0.2$ \\
            $a_{7}=1$ & $a_{8}=0$ & $a_{9}=0$ \\
            \bottomrule
        \end{tabular}
    \end{center}
    \caption{Parameter values for the bispectrum fitting formula used in~\cite{Scoccimarro:1997st}.}
    \label{tab:scocc_fit} 
\end{table}
\begin{table}[h]
    \begin{center} 
        \begin{tabular}{l|l|l} 
            \toprule
            \midrule
            $a_{1}=0.484$ & $a_{2}=3.740$ & $a_{3}=-0.849$ \\
            $a_{4}=0.392$ & $a_{5}=1.013$ & $a_{6}=-0.575$ \\
            $a_{7}=0.128$ & $a_{8}=-0.722$ & $a_{9}=-0.926$ \\
            \bottomrule
        \end{tabular}
    \end{center}
    \caption{Parameter values for the bispectrum fitting formula from~\cite{Gil-Marin:2011jtv}, used in this work.}
    \label{tab:hgm_fit} 
\end{table}

In order to solve Equation~(\ref{eq:xi_min_chi}), and get a functional form for $M\left(k,z\right)$, one can define an effective amplitude, $Q_{\rm eff}\left(r\right)$, as:
\begin{align}\label{eq:Qeff_chi2}
    Q_{\rm eff}\left(r\right)\left[\chi\left(r\right)\right]^{2} = \int & \frac{\dd^{3}k_{1}\dd^{3}k_{2}}{\left(2\pi\right)^{3}}\left[1-\left(\vec{\hat{k}}_{1}\cdot \vec{\hat{k}}_{2}\right)^{2}\right] B\left(\vec{k}_{1},\vec{k}_{2}\right) \nonumber\\
    & \times M\left(\vec{k}_{1}\right)M\left(\vec{k}_{2}\right)e^{-i \vec{k}_{12}\cdot \vec{r}},
\end{align}
which converts Equation~(\ref{eq:xi_min_chi}) to a second order algebraic equation for $\chi\left(r\right)$, which can then be solved for $M\left(k\right)$:
\begin{equation}\label{eq:M_iter}
    M\left(k,z\right) = \int \dd^{3}r~e^{i \vec{k} \cdot \vec{r}} \frac{\sqrt{1+1.5s^{3}(z)Q_{\rm eff}\left(r,z\right)\xi\left(r,z\right)}}{0.75s^{3}(z)Q_{\rm eff}\left(r,z\right)P\left(k,z\right)}.
\end{equation}
It is easy to see that Equation~(\ref{eq:M_iter}) still satisfies the asymptotic behavior in Equation~(\ref{eq:M_asymp}), and with this expression~\cite{Scoccimarro:2009eu} proposes the following iteration process to find an expression for $M\left(k\right)$ that is closer to its true solution Equation~(\ref{eq:M_full}):
\begin{itemize}
    \item[i)] One starts with the naive expression for the spherically symmetric case Equation~(\ref{eq:Gamma_Psi_Delta}) for the modified propagator, and evaluate $M_{\rm ini}\left(k\right)$ in Equation~(\ref{eq:Gamma_Psi_M}) as: 
    \begin{equation}\label{eq:M_ini}
        M_{\rm ini}\left(k,z\right) = \frac{1}{\sqrt{1+s^{3}\Delta^{ \mathrm{hf}}\left(k,z\right)}},
    \end{equation}
    where $\Delta^{\mathrm{hf}}\left(k,z\right)$ is the square root of the non-linear dimensionless power spectrum computed using \texttt{halofit}, but with the first order growth factor from the modified gravity theory considered.
    \item[ii)] With this $M_{\rm ini}\left(k\right)$, one evaluates $Q_{\rm eff}\left(r\right)$ by computing the integral in Equation~(\ref{eq:Qeff_chi2}), and find a new $M_{\rm new}\left(k\right)$ from Equation~(\ref{eq:M_iter});
    \item[iii)] With this new solution, $M_{\rm new}\left(k\right)$, one repeats the process in step (ii), i.e., one evaluates Equation~(\ref{eq:Qeff_chi2}) and Equation~(\ref{eq:M_iter}), until enough iterations are needed in order to achieve convergence for a certain $M_{\rm final}\left(k\right)$.
\end{itemize}
It is worth stressing that starting from an initial guess using \texttt{halofit} should not affect the final result too much, as an incorrect amplitude is irrelevant to us, only the asymptotic decay of $\Delta\left(k\right)$ is important, i.e., the shape of the dimensionless power spectrum. One can also infer from Equation~(\ref{eq:Qeff_chi2}), that, incorrectly evaluating the amplitude of $M_{\rm ini}$ is consistently cancelled on each side, as we have the same term $M_{\rm ini}^{2}\left(k\right)$ appearing in both sides of the Equation, explicitly on the right hand side, and implicitly inside $\chi\left(r\right)$ on the left hand side. We have tested different initial guesses, $M_{\rm ini}\left(k\right)$, by multiplying $\Delta^{\rm hf}\left(k\right)$ by different orders of magnitude, and present these results in Appendix~\ref{sec:AppA}. We can conclude that the final solution is independent of the amplitude of the initial guess, $M_{\rm ini}\left(k\right)$, as hoped for.

While this prescription seems consistent and simple enough, computing it in a fast and accurate way is not so straightforward. This can be seen by the need to perform one computationally expensive integral, Equation~(\ref{eq:Qeff_chi2}). In order to compute this integral we begin by expanding the term in front of the exponential,
\begin{align}\label{eq:B_tilde}
    \tilde{B}\left(k_{1},k_{2}, \mu_{12}\right) = \left(1-\mu_{12}^{2}\right) B\left(k_{1},k_{2}, \mu_{12}\right) M\left(k_{1}\right)M\left(k_{2}\right),
\end{align}
in Equation~(\ref{eq:Qeff_chi2}) in terms of Legendre polynomials:
\begin{equation}
     \tilde{B}\left(k_{1},k_{2}, \mu_{12}\right) = \sum_{\ell_{12}} \tilde{B}\left(k_{1},k_{2}\right)\mathcal{L}_{\ell_{12}}\left(\mu_{12}\right),
\end{equation}
as well as the exponential terms in plane waves:
\begin{align}
    e^{i\vec{k_{1}}\cdot \vec{r}} &= \sum_{\ell_{1}}\left(2\ell_{1}+1\right)i^{\ell_{1}}j_{\ell_{1}}\left(k_{1}r\right)\mathcal{L}_{\ell_{1}}\left(\mu_{1}\right),\\
    e^{i\vec{k_{2}}\cdot \vec{r}} &= \sum_{\ell_{2}}\left(2\ell_{2}+1\right)i^{\ell_{2}}j_{\ell_{2}}\left(k_{2}r\right)\mathcal{L}_{\ell_{2}}\left(\mu_{2}\right),
\end{align}
where $\mu_{12}$ is the cosine of the angle between the wave-numbers $\vec{k}_{1}$ and $\vec{k}_{2}$, $\mu_{1}$ and $\mu_{2}$ is the cosine between $\vec{r}$ and the wave-numbers $\vec{k}_{1}$ and $\vec{k}_{2}$ respectively and $j_\ell$ is the speherical Bessel function. If we then express the Legendre polynomials in terms of spherical harmonics, and use the orthogonality relations described in Appendix D1 of~\cite{Umeh:2020zhp}, we can perform the angular integrations that will leave us with $\ell_{1}=\ell_{2}=\ell_{12}$ and the integral is then rewritten as:
\begin{equation}\label{eq:Qeff_chi2_decomp}
    Q_{\rm eff}\left(r\right)\left[\chi\left(r\right)\right]^{2} = \sum_{\ell_{12}}i^{2\ell_{12}}\int\frac{\dd k_{1}}{k_{1}}\frac{\dd k_{2}}{k_{2}}\Tilde{\Delta}^{\Tilde{B}}_{\ell_{12}} \left(k_{1},k_{2}\right)j_{\ell_{12}}\left(k_{1}r\right)j_{\ell_{12}}\left(k_{2}r\right),
\end{equation}
with
\begin{equation}\label{eq:Delta_tilde_B}
    \Tilde{\Delta}^{\Tilde{B}}_{\ell_{12}}\left(k_{1},k_{2}\right)  = \frac{k_{1}^{3}k_{2}^{3}\Tilde{B}_{\ell_{12}}\left(k_{1}, k_{2}\right)}{\left(2\pi^{2}\right)^{2}},
\end{equation}
where we have defined
\begin{align}\label{eq:B_tilde_recons}
    &\Tilde{B}_{\ell_{12}}\left(k_{1}, k_{2}\right) = \frac{2\ell_{12}+1}{2}  
    \int_{-1}^{1}\dd \mu_{12} \tilde{B}\left(k_{1},k_{2}, \mu_{12}\right) \mathcal{L}_{\ell_{12}}\left(\mu_{12}\right).
\end{align}

\begin{figure}[h] 
\centering
\includegraphics[width=1.\textwidth]{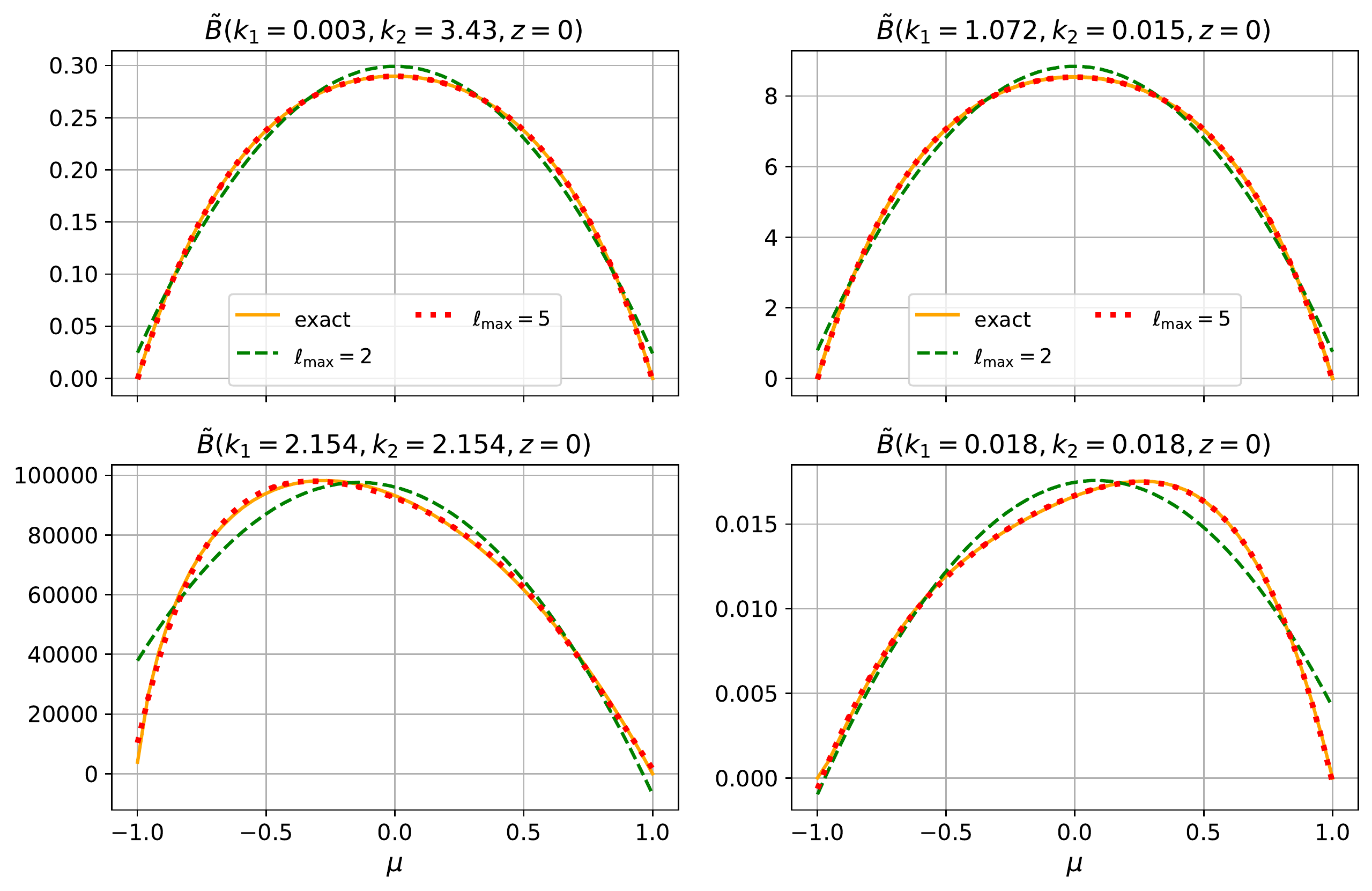}
\caption{Reconstruction of the quantity given in Equation~(\ref{eq:B_tilde}) using Legendre polynomials until different maximum multipoles. Solid orange line is the exact quantity, dashed green is the reconstructed quantity truncating at $\ell_{\rm max}=2$ and dotted red lines truncating at $\ell_{\rm max}=5$.}
\label{fig:bspt_recons}
\end{figure}

The integral on the right hand side of Equation~(\ref{eq:Qeff_chi2_decomp}) can be solved using the methods described in~\cite{Fang:2020vhc,Umeh:2020zhp}, which extend the \texttt{FFTLog} method to two-dimensions, i.e., \texttt{FFTLog2D}, and provides a publicly available code to do this calculation\footnote{\hyperlink{https://github.com/xfangcosmo/2DFFTLog}{https://github.com/xfangcosmo/2DFFTLog}}. While Equation~(\ref{eq:Qeff_chi2_decomp}) can now be evaluated quickly, it is a series expansion, and needs to be truncated at a maximum multipole, $\ell_{12}^{\rm max}$, once a good agreement with the exact expression is found. It turns out that there is no need to go to very high multipoles in order to reconstruct the exact expression in Equation~(\ref{eq:B_tilde}), and we show in Figure~\ref{fig:bspt_recons} the reconstructed quantity Equation~(\ref{eq:B_tilde_recons}) for four different triangle configurations using two different maximum values for $\ell_{12}$, $\ell_{12}^{\rm max}=3$ and $\ell_{12}^{\rm max}=5$.
%%%%%%%%%%%%%%%%
%%%%%%%%%%%%%%%%
As we can see, $\ell_{12}^{\max} = 5$ already gives a very good agreement in the four triangle configurations we show. Therefore, in the results presented in subsequent sections we will always use this value as the maximum multipole to compute the reconstructed $\tilde{B}\left(k_{1},k_{2}, \mu_{12}\right)$. Another important remark to note is that even though we formally expand Equation~(\ref{eq:B_tilde_recons}), the quantity we are interested in computing is actually $Q_{\rm eff}$, which characterizes the average response of the Vainshtein mechanism in the growth of structure. The effects of screening are only present at small scales, i.e. for small values of $r$. In this way, by adding more multipoles to the expansion will mainly affect the reconstruction at large scales, where $r \gg R_{V}$, and it is where we know that the linear theory is the correct description of the system, and the asymptotic behavior of the function $M(k)$ is known (see Equation~(\ref{eq:M_asymp})). Due to this, we believe that truncating the series at a not so large multipoles is enough to capture the non-linear effects we are interested in. The computation of the integral in Equation~(\ref{eq:M_iter}) is relatively simple, as we can just use a cosmological code that performs 1-dimensional FFTLog algorithm, such as \texttt{nbodykit}~\cite{Hand:2017pqn}\footnote{\hyperlink{https://nbodykit.readthedocs.io/en/latest/}{https://nbodykit.readthedocs.io/en/latest/}} used in this work.

\section{Simulations}\label{sec:sims}
In this section we will discuss the simulations and their specifications used in this work. As previously mentioned, we will consider two modified gravity theories, nDGP and Cubic Galileon, and our results obtained by our COLA simulations that implement the screening approximation outlined in the previous section will be compared with full {\it N}-body simulation suites available in the literature for these two models. 

\begin{table}[h]
    \begin{center} 
        \begin{tabular}{l|l} 
            \toprule
            \midrule
            $\Omega_{\rm m0}$ & $0.281$ \\
            $\Omega_{\rm \Lambda}$ & $0.719$  \\
            $\Omega_{\rm b0}$ & $0.046$  \\
            $n_{\rm s}$ & $0.971$  \\
            $\sigma_{8}$ & $0.842$  \\
            $h$ & $0.697$  \\
            \bottomrule
        \end{tabular}
    \end{center}
    \caption{The cosmological parameters of the ELEPHANT simulation suite.}
    \label{tab:ndgp_cosmo} 
\end{table}

For the nDGP suite, we will use the so-called \texttt{ELEPHANT} suite, which was run using the \texttt{ECOSMOG} code~\cite{esmog}, an adaptive-mesh-refinement modified gravity {\it N}-body code that is built using the architecture of the GR {\it N}-body code \texttt{RAMSES}~\cite{Teyssier:2001cp}. In this suite, we have two cases of nDGP gravity: $r_{c}H_{0}=1.0$ and $r_{c}H_{0}=5.0$, where the former exhibits a stronger effect of modified gravity on structure formation.
%, while the latter gives us a smaller deviation from the GR predictions. 
The cosmological parameters used are shown in Table~\ref{tab:ndgp_cosmo}. These simulations were initiated at redshift $z_{\rm ini}=49$, and the initial conditions were generated using the \texttt{MPGraphic} code~\cite{Prunet:2008fv}, that creates particle displacements using the Zel'dovich approximation~\cite{Zeldovich:1969sb}, i.e., 1LPT, and the linear input matter power spectra was generated using \texttt{CAMB} at $z=0$ using the cosmological parameters in Table~\ref{tab:ndgp_cosmo}. The specifications for these simulations are shown on the left side of Table~\ref{tab:NB_specs}.

\begin{table}[!htb]
    \begin{minipage}{.5\linewidth}
      %\caption{}
      \vspace{0.2cm}
      \texttt{ELEPHANT}
      \centering
        \begin{tabular}{ll}
        \toprule
        %ss
        \midrule
        %Models & N1, N5 \\
        Realisations & 5 \\
        Box size & 1024  \\
        $N_{\mathrm{part}}$ & $1024^3$ \\
        Domain grid & $1024^3$ \\
        Refinement criterion & $8$ \\
        Initial conditions & Zel'dovich, $z=49$ \\
        \bottomrule						
        \end{tabular}
    \end{minipage}%
    \begin{minipage}{.5\linewidth}
      \vspace{0.2cm}
    \texttt{Cubic Galileon}
      \centering
        %\caption{}
        \begin{tabular}{ll}
        \toprule
        \midrule
       %Models & Cubic Galileon \\
        Realisations & 5 \\
        Box size & $400~(200)$  \\
        $N_{\mathrm{part}}$ & $512^3$ \\
        Domain grid & $1024^3$ \\
        Refinement criterion & $6~(8)$ \\
        Initial conditions & Zel'dovich, $z=49$ \\
        \bottomrule
        \end{tabular}
    \end{minipage} 
    \caption{Specifications of the {\it N}-body simulations used in this work. \textbf{Left:} nDGP simulations. \textbf{Right:} Cubic Galileon. All simulations were run using the \texttt{ECOSMOG} code, for the Cubic Galileon case the authors have run simulations for two boxsizes, $400$ Mpc$/h$ and $200$ Mpc$/h$. However, in this work we will only use the first box.}
    \label{tab:NB_specs}
\end{table}

For the case of the Cubic Galileons model, we have used the {\it N}-body simulations from~\cite{Barreira:2013eea}, that were also ran using the \texttt{ECOSMOG} code, and the initial conditions were also generated with the same code as for the \texttt{ELEPHANT} suite. 

\begin{table}[h]
    \begin{center} 
        \begin{tabular}{l|l} 
            \toprule
            \midrule
            $\Omega_{\rm cdm 0}h^{2}$ & $0.1274$ \\
            $\Omega_{\rm b 0}h^{2}$ & $0.02196$  \\
            $n_{\rm s}$ & $0.953$  \\
            $\ln\left(10^{10}A_{s}\right)$ & $3.154$  \\
            $h$ & $0.7307$  \\
            $\ln\left(\Omega_{\phi_{i}}/\Omega_{\rm m,i}\right)$ & $-4.22$ \\
            $c_{2}/c_{3}^{2/3}$ & $-5.378$ \\
            $c_{3}$ & $10$ \\
            \bottomrule
        \end{tabular}
    \end{center}
    \caption{The cosmological and Cubic Galileon parameters used in \cite{Barreira:2013eea}.}
    \label{tab:cugal_cosmo} 
\end{table}
In Table~\ref{tab:cugal_cosmo} we show the cosmological and Cubic Galileon parameters used in~\cite{Barreira:2013eea} for their {\it N}-body simulations. In order to connect the last two rows of Table~\ref{tab:cugal_cosmo}, we show the identification between the Cubic Galileon model and the Horndeski action in Equation~(\ref{eq:actionhorn}):
\begin{subequations}\label{eq:cugal_Gis}
\begin{align}
    &G_{4}\left(\phi, X\right) = \frac{1}{2}, \\
    &G_{3}\left(\phi, X\right) = \frac{c_{3}}{M^{3}}\dot{\phi}^{2},\\
    &K\left(\phi, X\right) = \frac{c_{2}}{2}\dot{\phi}^{2}.
\end{align}
\end{subequations}
To further understand how the Cubic Galileon model is  specified from the values of $c_{2}$ and $c_{3}$, as well as the initial fractional energy density of the scalar field $\Omega_{\phi_{i}}$ we refer to References~\cite{Barreira:2012kk,Barreira:2013eea}. 
%One last remark concerning this suite of {\it N}-body simulations, is with respect to the theories the authors have ran simulations. 
In order to understand and highlight the impact of the non-linear clustering introduced by the galileon scalar field, as well as the Vainshtein screening effects at small scales, the simulations were run for three different theories, shown in Table~\ref{tab:cugal_sims}.
\begin{table}[h]
    \begin{center} 
        \begin{tabular}{lcccc} 
            \toprule
            Model & Background  &
            Linear Growth &
            Screening \\ 
            \midrule
            QCDM & Cubic Galileon & GR & No need \\
            Linear Cubic Galileon & Cubic Galileon & Cubic Galileon & No \\
            Full Cubic Galileon & Cubic Galileon & Cubic Galileon & Yes \\
            \bottomrule						
        \end{tabular}
    \end{center}
    \caption{Cubic Galileon models considered in \cite{Barreira:2013eea}.}
    \label{tab:cugal_sims} 
\end{table}
The settings for each simulation is shown on the right of Table~\ref{tab:NB_specs}, where the number of particles in all simulations were fixed to $512$ per dimension, and two different boxsizes was considered: $400~\mathrm{Mpc}/h$ and $200~\mathrm{Mpc}/h$. For each model in Table~\ref{tab:cugal_sims}, a total of $10$ simulations were run, $5$ per different boxsize. The QCDM model is defined as having the cubic galileon expansion history in the background, but its linear and non-linear growth follows GR. The Linear Cubic Galileon case, has the cubic galileon expansion history and linear growth, however, there is no screening to shield the fifth force introduced by the galileon scalar field. The last model is the Full Cubic Galileon model, where background, linear growth, non-linear growth and screening are present, and this represent the complete implementation that we want to compare with.

In this work, we have also run simulations using the COLA method, which is an approximate {\it N}-body technique to quickly generate non-linear realizations of the matter density field. COLA combines 2LPT to accurately describe the large and intermediate scales behavior of the density field, while implementing a Particle-Mesh (PM) algorithm~\cite{PM1988} to evolve particles at small scales. Using this prescription, COLA is able to compute the matter power spectrum with fewer time-steps than usual {\it N}-body simulations. Since both of the full {\it N}-body suites used in this work have initial redshift $z_{\rm ini}=49$, we also start our COLA simulations with the same redshift, and, the time-stepping choice used in our runs are shown in Table~\ref{tab:COLA_ndgp_ts} and Table~\ref{tab:COLA_cugal_ts} for the nDGP models and Cubic Galileon respectively. In both time-stepping choices, we have chosen a time-resolution of $\dd a \approx 0.02$, and each time-step interval is linearly spaced in time.
\begin{table}[h]
    \begin{center} 
        \begin{tabular}{lc} 
            \toprule
            Redshift interval & Number of time-steps \\
            \midrule
            $49 \to 1.628$  & $17$ \\
            $1.628 \to 1.432$  & $2$ \\
            $1.432 \to 1.172$  & $2$ \\
            $1.172 \to 1.017$  & $2$ \\
            $1.017 \to 0.507$  & $9$ \\
            $0.507 \to 0$  & $18$ \\
            \bottomrule
        \end{tabular}
    \end{center}
    \caption{Number of time-steps intervals for all COLA simulations for the nDGP models considered in this work.}
    \label{tab:COLA_ndgp_ts} 
\end{table}

\begin{table}[h]
    \begin{center} 
        \begin{tabular}{lc} 
            \toprule
            Redshift interval & Number of time-steps \\
            \midrule
            $49 \to 0.667$  & $30$ \\
            $0.667 \to 0.25$  & $10$ \\
            $0.25 \to 0$  & $10$ \\
            \bottomrule
        \end{tabular}
    \end{center}
    \caption{Number of time-steps intervals for all COLA simulations for the Cubic Galileon model considered in this work.}
    \label{tab:COLA_cugal_ts} 
\end{table}

For the generation of the initial conditions (IC) of our COLA simulations, we proceeded in two different ways, one that was used to present the results in the main text in Section~\ref{sec:results}, and another one in which the comparisons are left for Appendix~\ref{sec:AppB}. In the former approach ICs were generated from an initial random Gaussian field with a fixed amplitude, and the initial particle displacements computed using 2LPT. For the latter, we read particle positions directly from the snapshot at $z=49$ used in one of {\it N}-body simulations and reconstruct the displacement field using the Zel'dovich approximation (1LPT). The main difference between the two is that in the second IC generation much of the noise coming from cosmic variance effects is cancelled out when comparing the two power spectra directly. However, as shown and discussed in Appendix~\ref{sec:AppB}, this method still introduces other difficulties that keep us from reproducing exactly the results from the {\it N}-body simulations, namely, the displacement fields in {\it N}-body simulations were computed using only 1LPT, while our COLA simulations use 2LPT to compute the same displacement fields. This ends up introducing transient effects~\cite{Crocce:2006ve} at intermediate redshifts in the power spectra, pushing the amplitude of our COLA simulations a few percent off from the {\it N}-body results. These effects are largely cancelled out if we take the ratio of the power spectrum in a modified gravity model and a corresponding GR model.

The settings of our COLA simulations are shown in Table~\ref{tab:COLA_specs} for each modified gravity theory. Specifically for the Cubic Galileon case, we run another set of COLA simulations with a more refined grid (increased force resolution), which allows us to test the convergence of our results, and we show this comparison in Appendix~\ref{sec:AppB}.
\begin{table}[h]
    \begin{center} 
        \begin{tabular}{lcccc} 
            \toprule
            Sim. & Vol.  &
            Num. of particles &
            Num. of grids &
            Num. of time steps \\ 
            \midrule
            COLA nDGP & $512^3$ 
            & $512^3$ & $(2N_{\rm p})^3$= $1024^3$ & 50 \\
            COLA Cubic-Gal. & $400^3$ 
            & $512^3$ & $(2N_{\rm p})^3$=$1024^3$ & 50 \\
            COLA Cubic-Gal. & $400^3$ 
            & $512^3$ & $(3N_{\rm p})^3$=$1536^3$ & 50 \\
            \bottomrule						
        \end{tabular}
    \end{center}
    \caption{COLA simulation specifications for the two modified theories of gravity considered in this work. The volume is in the unit of {\rm Mpc}$^{3}$/$h^{3}$.}
    \label{tab:COLA_specs} 
\end{table}

\section{Results}\label{sec:results}

In this section we will show the results for two theories of modified gravity: nDGP (normal branch DGP) and the Cubic Galileon. 

\subsection{nDGP}\label{sec:results_ndgp}

\begin{figure}[h] 
\centering
\includegraphics[width=1.\textwidth]{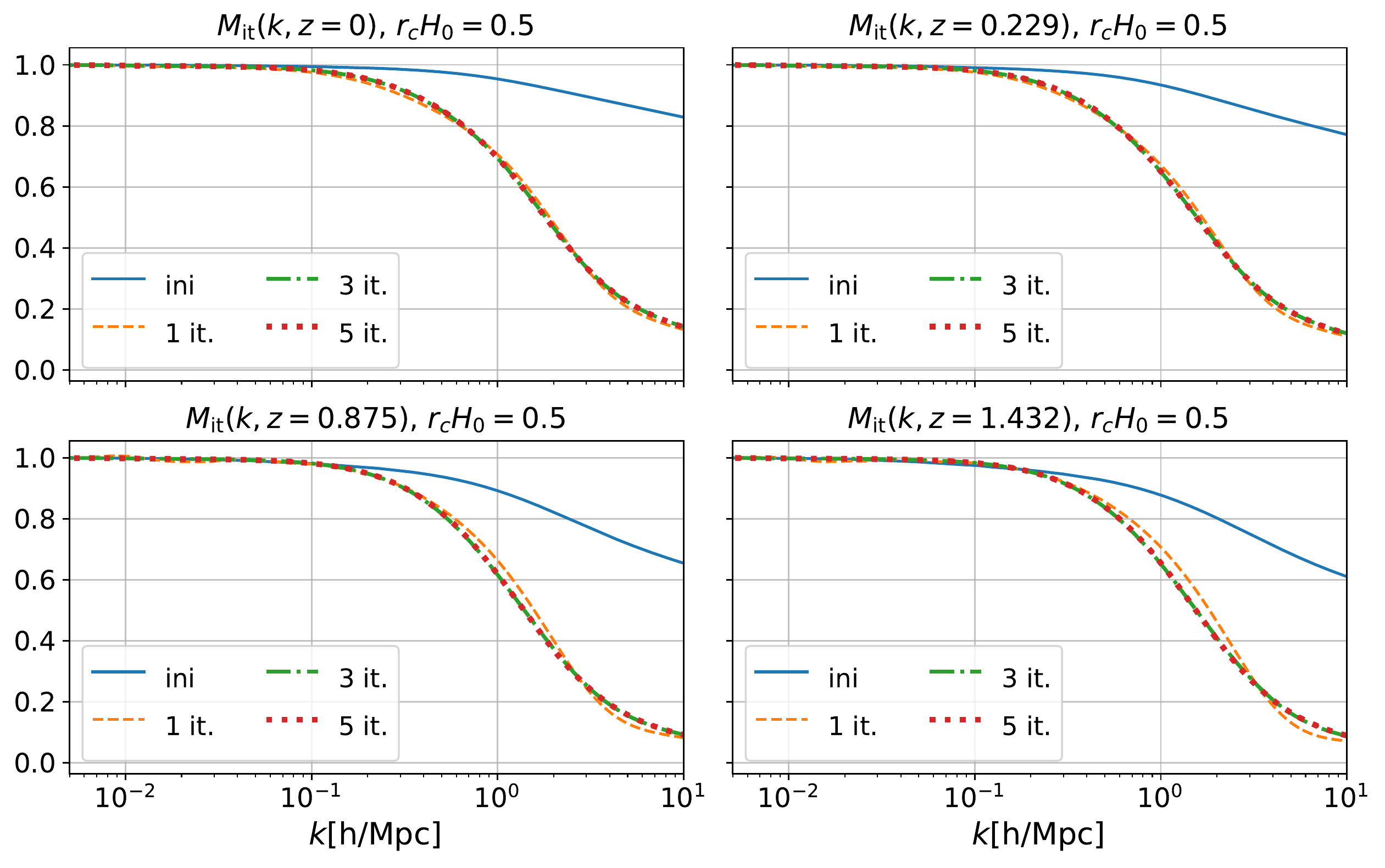}
\caption{Different iteration values for $M\left(k,z\right)$ at different redshifts as a function of wave-number for the nDGP gravity model with $r_{c}H_{0}=0.5$, which represents the biggest deviation from GR. The solid blue curves represent the initial guess, the dashed orange the quantity after one iteration, dotted-dashed green after 3 iterations and dotted red 5 iterations.}
\label{fig:M_conv_05}
\end{figure}

\begin{figure}[h] 
\centering
\includegraphics[width=1.\textwidth]{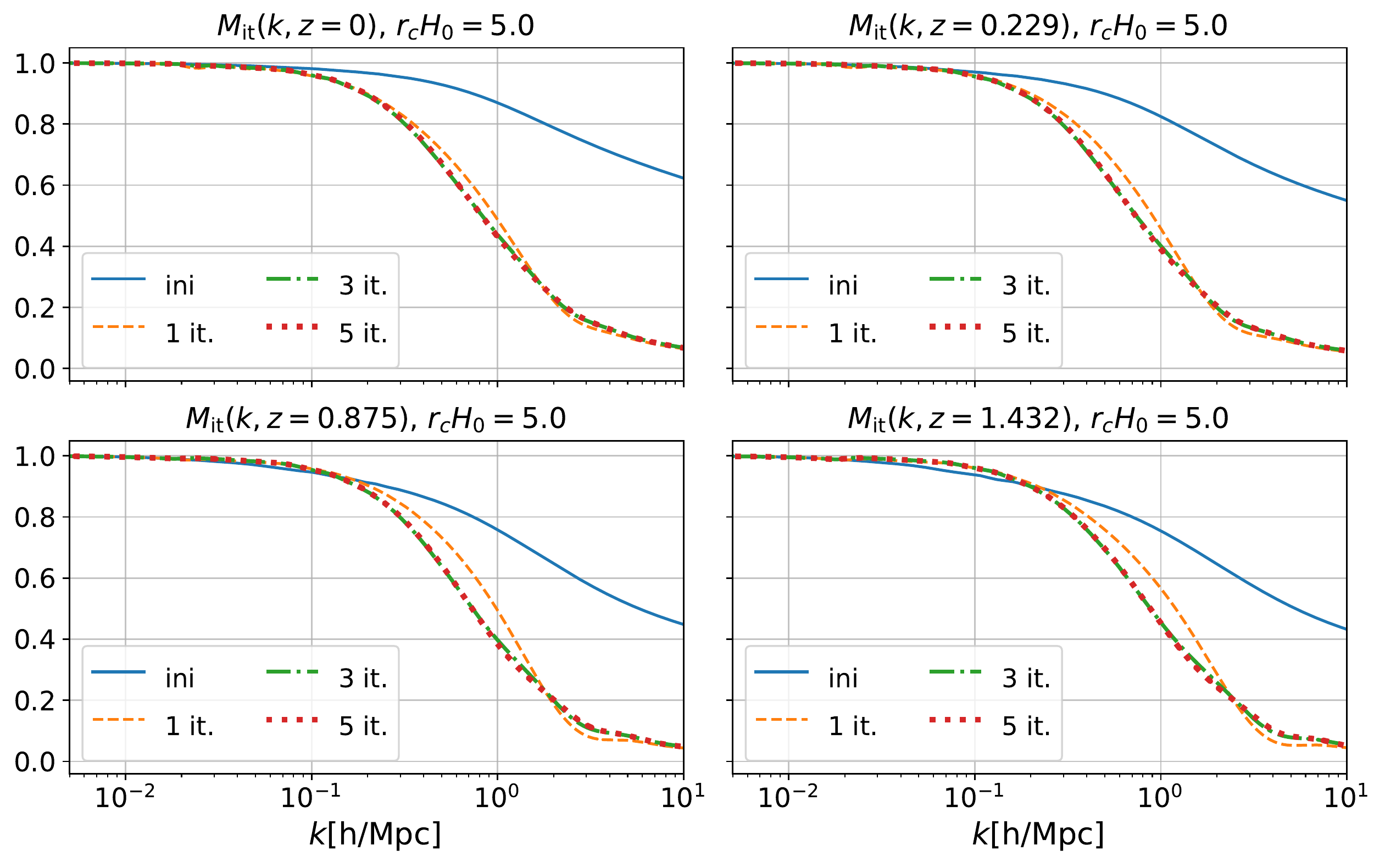}
\caption{Different iteration values for $M\left(k,z\right)$ at different redshifts as a function of wave-number for the nDGP gravity model with $r_{c}H_{0}=5.0$, which represents the smallest deviation from GR. The solid blue curves represent the initial guess, the dashed orange the quantity after one iteration, dotted-dashed green after 3 iterations and dotted red 5 iterations.}
\label{fig:M_conv_5}
\end{figure}

\begin{figure}[h] 
\centering
\includegraphics[width=1.\textwidth]{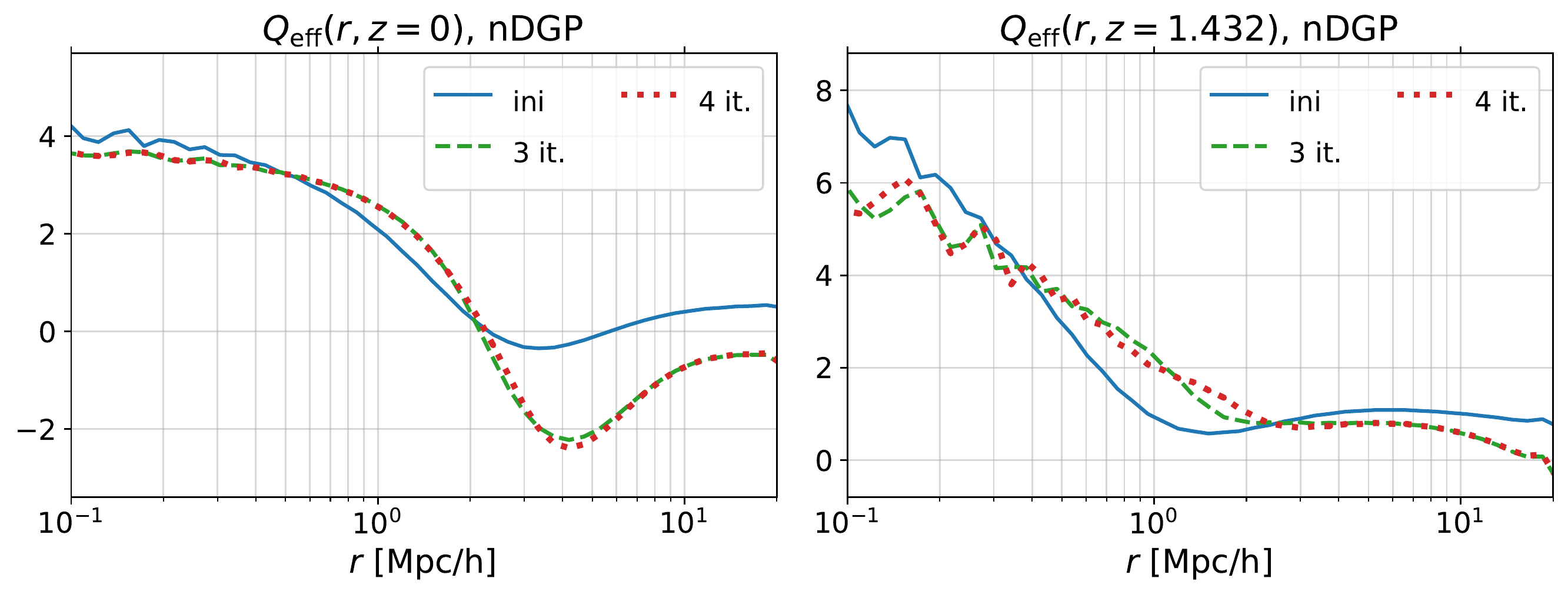}
\caption{$Q_{\rm eff}(r)$ as a function of scale at two redshifts, left $z=0$ and right $z=1.432$. We can see tat the curves also overlap after some iterations.}
\label{fig:Q_eff}
\end{figure}

\begin{figure}[h] 
\centering
\includegraphics[width=1.\textwidth]{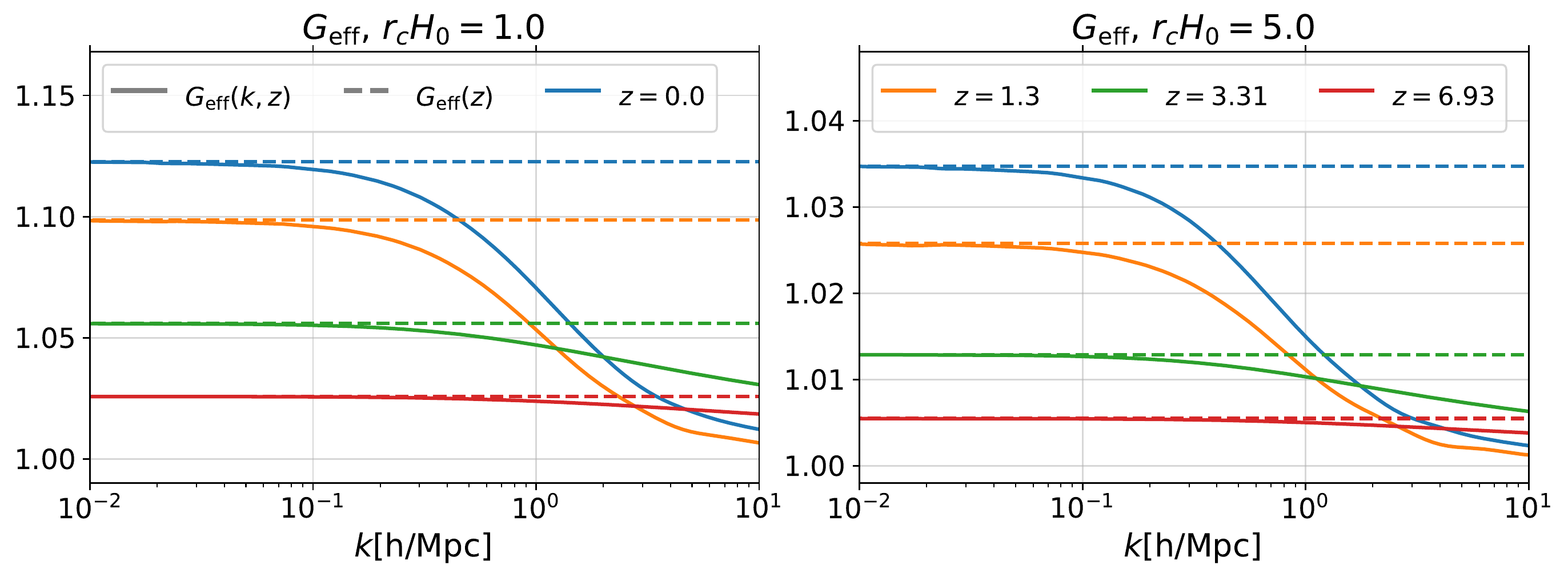}
\caption{\textbf{Left:} Linear $G_{\rm eff}$ (solid lines) and non-linear $G_{\rm eff}$ (dashed lines) for nDGP gravity model $r_{c}H_{0}=1.0$ at four different redshifts. \textbf{Right:} The same quantity but for nDGP gravity model $r_{c}H_{0}=5.0$.}
\label{fig:Geff_lin_nl_ndgp}
\end{figure}

\begin{figure}[h] 
\centering
\includegraphics[width=1.\textwidth]{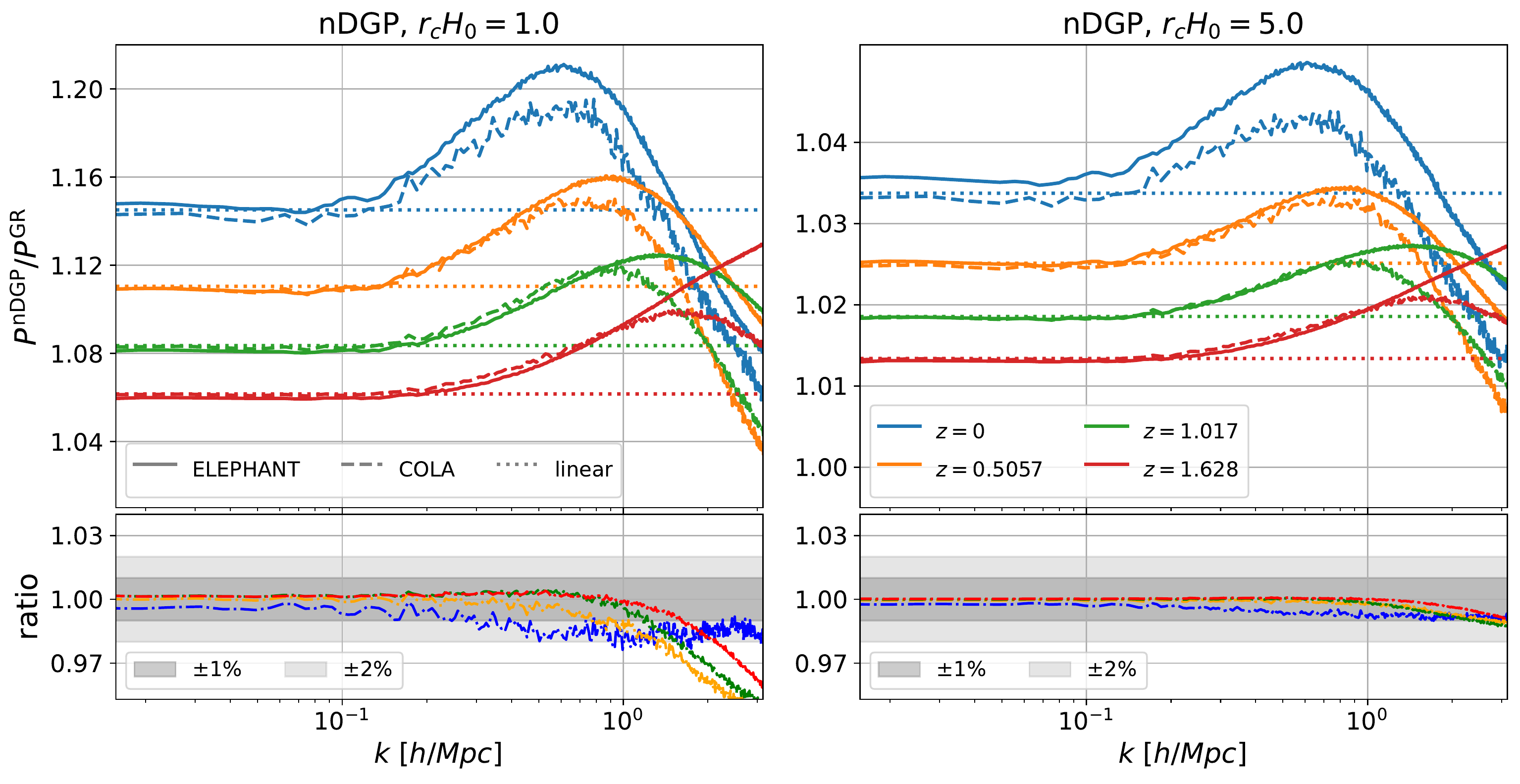}
\caption{Ratio of the power spectrum between nDGP gravity and GR from the \texttt{ELEPHANT} suite (solid lines), COLA (dashed lines) and linear prediction, at four different redshifts, for two models of nDGP gravity $r_{c}H_{0}=1.0$ (left) and $r_{c}H_{0}=5.0$(right).}
\label{fig:ndgp_GR_ratio}
\end{figure}

We begin by showing results for the nDGP theory, where we ran COLA simulations for a total of three different parameter choices: $r_{c}H_{0}=0.5, 1.0$ and $5.0$, where the smaller the value the greater the deviation with respect to GR (see right hand side plot of Figure~\ref{fig:ndgp_bg}). In order to compare with the \texttt{ELEPHANT} suite, we fixed the cosmological parameters to the same values used in the suite, shown in Table~\ref{tab:ndgp_cosmo}. We then perform the steps outlined in the previous section to determine $M\left(k,z\right)$ iteratively. As already mentioned, the non-linear fitting formula to the bispectrum proposed by \cite{Gil-Marin:2011jtv}, Equation~(\ref{eq:B_fit}) together with Equation~(\ref{eq:abc_funct_fit}), is only valid for redshifts smaller than $z_{\rm fit} = 1.5$, therefore, the iterative computation to find a convergent $M_{\rm final}\left(k,z\right)$ is performed only up to this limiting redshift. In Figures~\ref{fig:M_conv_05} and \ref{fig:M_conv_5} we show $M\left(k,z\right)$ at four different redshifts, for two different types of nDGP gravity, $r_{c}H_{0}=0.5$ and $r_{c}H_{0}=5.0$, the greatest and smallest modification with respect to GR respectively and, for four different iteration steps: the initial guess (ini), with $1$ iteration (1 it.), with $3$ iterations (3 it.) and with $5$ iterations (5 it.). We can see that we have a great improvement already after one iteration, as the curve is pushed down considerably at scales $k > 0.1$ $h/$Mpc, where screening effects become relevant.
%%%%%%%%%%%%%%%%
%%%%%%%%%%%%%%%%
%%%%%%%%%%%%%%%%
%%%%%%%%%%%%%%%%
After only a few iterations the solution are seen to have converged: the 3 iterations curve, green dot-dashed, and the 5 iterations one, red dotted, are overlapping on all scales of interest. This shows that after a few iterations we are already able to find a convergent solution, and one does not need to perform the iteration process many times, thus, saving time. Analogously, we show in Figure~\ref{fig:Q_eff} the quantity $Q_{\rm eff}$ for some iteration steps at two redshifts, $z=0$ and $z=1.432$ the limiting redshifts for which our procedure is performed, for nDGP gravity $r_{c}H_{0}=1.0$. A comparison with the top plot of Figure~10 in~\cite{Scoccimarro:2009eu} shows that the curve of our initial guess is in agreement with its counterpart on the cited reference. The behavior of the curves coming from the iteration steps, however, is different than the one in the Figure~10 of~\cite{Scoccimarro:2009eu}. This is due to the fact that in the original Reference~\cite{Scoccimarro:2009eu}, the DGP gravity model used in question is the self-accelerating branch of the action in Equation~(\ref{eq:dgp_action}). This theory has a very different background history, as one observes ``degravitation'' at the linear level, i.e., the growth is suppressed instead of enhanced. Therefore, one can expect that the behavior of the iterative solutions of $Q_{\rm eff}\left(r,z\right)$ to exhibit an opposite behavior. The function $M_{\rm it}\left(k,z\right)$, however, must follow the same behavior of being suppressed after its initial guess, as this function essentially must capture the transition between linear scales (unscreened) and non-linear scales (screened).
%%%%%%%%%%%%%%%%%%%%%
%%%%%%%%%%%%%%%%%%%%%
Once we have a convergent solution for $M_{\rm final}\left(k,z\right)$ at the redshifts in which the non-linear bispectrum fitting formula is valid, we can compute $G_{\rm eff}\left(k,z_{\rm fit}\right)$ following Equation~(\ref{eq:Geff_of_k_z}). Our COLA simulations, nevertheless, are initiated at $z_{\rm ini}=49$, while the $M_{\rm final}\left(k,z\right)$ we computed go up to $z_{\rm fit}=1.5$. To be consistent with the iteration formalism, we have decided to perform the same iteration process as before for higher redshifts, but, instead of using the non-linear fitting formula for the bispectrum, Equation~(\ref{eq:B_fit}), we use its tree-level formula~\cite{Bernardeau:2001qr}:
\begin{align}\label{eq:B_tree}
    B_{\rm tree}\left(\vec{k}_{1}, \vec{k}_{2}, \vec{k}_{3}\right) = 2 F_{2}\left(\vec{k}_{1}, \vec{k}_{2}\right) &P\left(\vec{k}_{1}\right)P_{\rm lin}\left(\vec{k}_{2}\right) + 2 F_{2}\left(\vec{k}_{2}, \vec{k}_{3}\right) P_{\rm lin}\left(\vec{k}_{2}\right)P_{\rm lin}\left(\vec{k}_{3}\right) \nonumber\\
    &+ 2 F_{2}\left(\vec{k}_{3}, \vec{k}_{1}\right) P_{\rm lin}\left(\vec{k}_{3}\right)P_{\rm lin}\left(\vec{k}_{1}\right),
\end{align}
where we then use the linear matter power spectrum $P_{\rm lin}\left(k,z\right)$ computed using \hiclass. In Figure~\ref{fig:Geff_lin_nl_ndgp} we show the final solutions for $G_{\rm eff}\left(k,z\right)$, solid lines lines, and its linear counterpart, dashed lines, for two nDGP gravity models the \texttt{ELEPHANT} suite was run, $r_{c}H_{0}=1.0$ and $r_{c}H_{0}=5.0$.
%%%%%%%%%%%%%%%%
%%%%%%%%%%%%%%%%
Now that we have $G_{\rm eff}$ as a function of time and wave-number, we then feed the tabulated data function to our COLA simulations at each time-step, with the code now solving the modified Poisson equation:
\begin{equation}
    \nabla^{2}\Phi\left(k,z\right) = 4 \pi G_{\rm eff}\left(k,z\right)\rho_{\rm m}\delta_{\rm m}\left(k,z\right),
\end{equation}
and the screening is implemented in Fourier space consistently. As one can see, this equation can be solved efficiently with fast Fourier transforms, hence implying an overall speed-up when contrasted with the alternative of solving the equation of motion for the scalar field. It is important to note that there is no introduction of extra parameters for the computation of $G_{\rm eff}(k,z)$, and this method requires no fitting with full {\it N}-body simulations in order to find the correct behavior in which the coupling between matter and gravity transitions from its linear theory value to its screened one. This is a desirable feature when trying to implement screening mechanisms, as screening is a model-dependent effect, and, within this formalism, all model-dependency is encoded in the non-linear time-dependent function $s^{3}(z)$, displayed on the left plot of Figure~\ref{fig:ndgp_bg}. The validation plots we are interested in are shown in Figure~\ref{fig:ndgp_GR_ratio}, where the top plots show the ratio between the non-linear cold dark matter power spectrum in nDGP, $r_{c}H_{0}=1.0$ on the left and $r_{c}H_{0}=5.0$ on the right, and GR. The dashed lines are the COLA predictions, the solid ones the \texttt{ELEPHANT} suite ones, and each color corresponds to a different redshift. We also plot the linear theory prediction, as a consistency check at small values of the wave-number.
%%%%%%%%%%%%%%%%
%%%%%%%%%%%%%%%%
We can see that for both nDGP gravity models, the agreement between the two different methods, COLA with approximate screening and \texttt{ELEPHANT} suite solving the full non-linear equations, is below $2\%$ for all scales of interest for current and future stage-IV LSS surveys, i.e., up to $k\approx 1~h/\mathrm{Mpc}$. The comparison of the ratio between the non-linear power spectra of a given modified theory of gravity and GR is valuable for two reasons: firstly because it removes errors coming from resolution effects and sample variance, and secondly because one of the greatest interests for the analysis of the data coming from next generation galaxy surveys relies on emulating the boost between modified theories of gravity and General Relativity, which is precisely the ratio plotted. The boost is formally defined as:
\begin{equation}\label{eq:boost}
    B^{\rm MG}\left(k,z\right) = \frac{P_{\rm NL}^{\rm MG}\left(k,z\right)}{P_{\rm NL}^{\rm GR}\left(k,z\right)}.
\end{equation}
This quantity has been shown to be sufficiently cosmology-independent~\cite{Kaushal:2021hqv,Brando:2022gvg}, allowing us to generate only a small set of simulations scanning the modified gravity parameter space, and train emulators with it.

\subsection{Cubic Galileon}\label{sec:results_cugal}

\begin{figure}[h] 
\centering
\includegraphics[width=1.\textwidth]{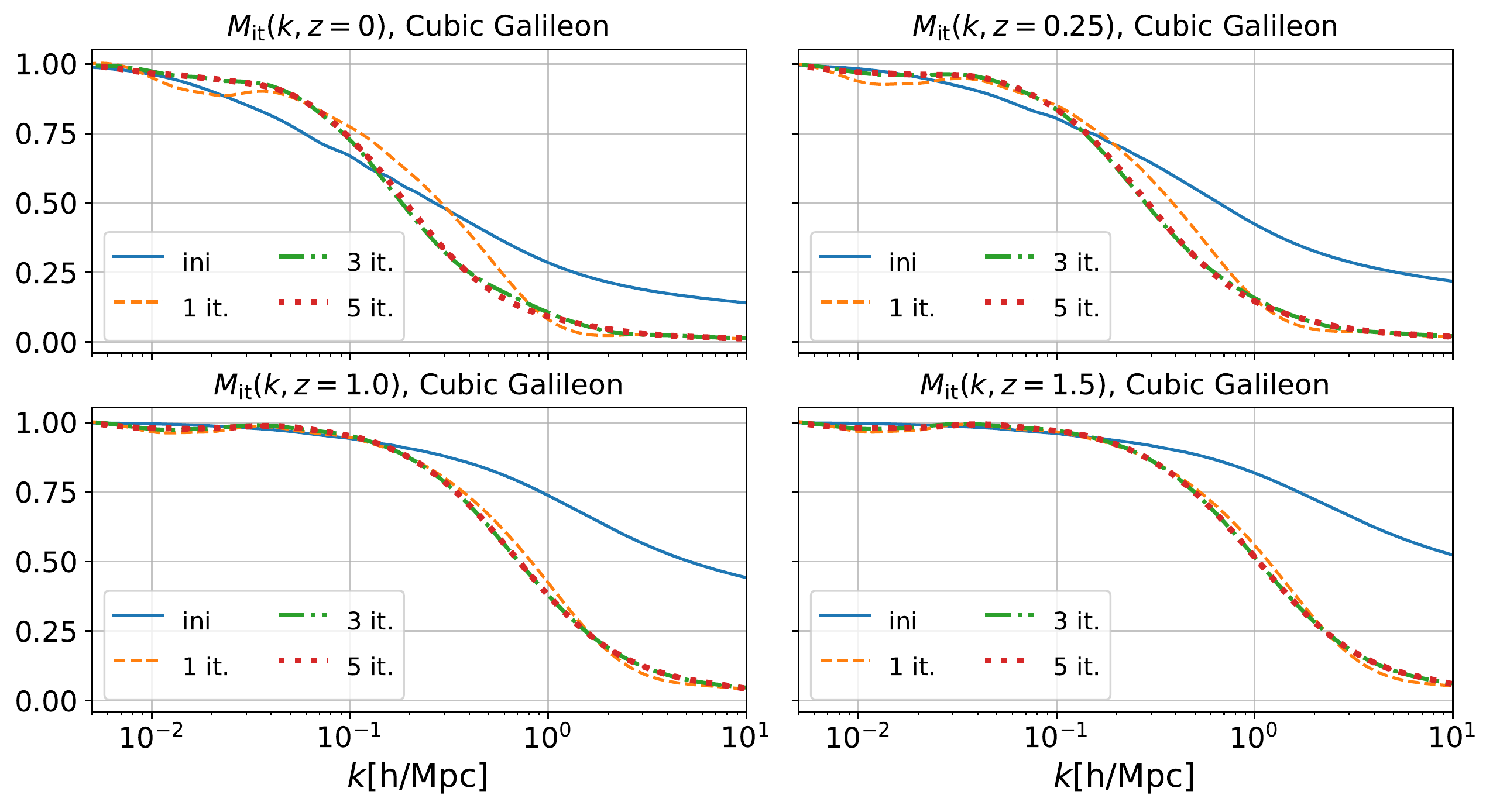}
\caption{Different iteration values for $M\left(k,z\right)$ at different redshifts as a function of wave-number for the Cubic Galileon model. The solid blue curves represent the initial guess, the dashed orange the quantity after one iteration, dotted-dashed green after 3 iterations and dotted red 5 iterations.}
\label{fig:M_conv_cugal}
\end{figure}

\begin{figure}[h] 
\centering
\includegraphics[width=0.75\textwidth]{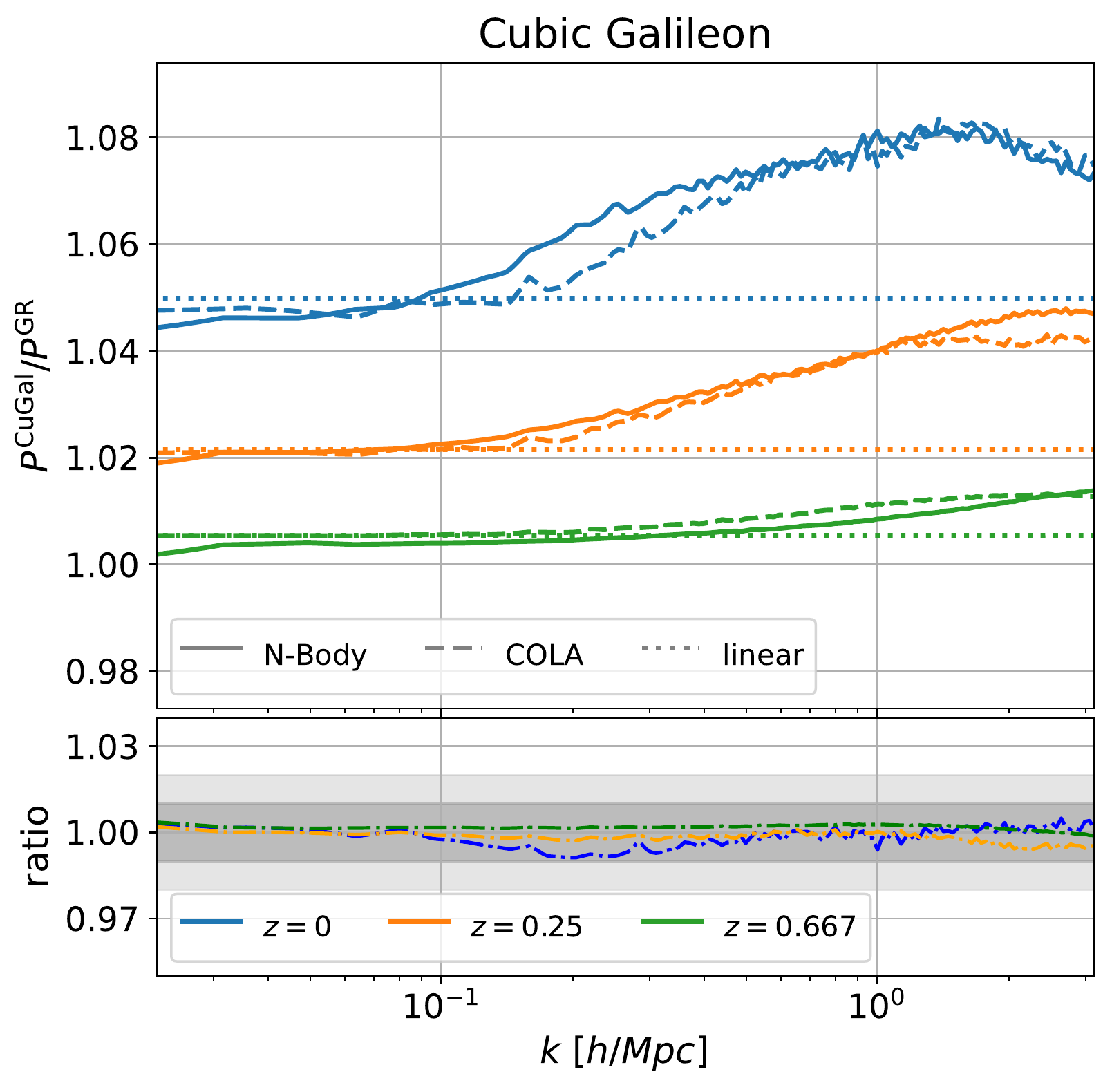}
\caption{Ratio of the power spectrum between the Cubic Galileon model and GR from {\it N}-body simulations (solid lines), COLA (dashed lines) and linear prediction, at three different redshifts.}
\label{fig:cugal_GR_ratio}
\end{figure}

Our next test case is the Cubic Galileon model, described by the action in Equation~(\ref{eq:actionhorn}) with the functions specified by Equation~(\ref{eq:cugal_Gis}). In order to consistently implement the Cubic Galileon model inside COLA, we need to implement the equations of motion for 2LPT. However, the Cubic Galileon model is a sub-case of Horndeski theories minimally coupled to the metric tensor Equation~(\ref{eq:MCHT}), therefore, in Appendix~\ref{sec:AppC} we present the general formulation for these theories that were implemented in COLA. The background quantities of this model are plotted in Figure~\ref{fig:ndgp_bg} in the dotted red curves. We can see that the behavior of the non-linear function Equation~(\ref{eq:s_3_bra}) is considerably different at late times when contrasted with the nDGP curves. This is due to the fact that the background in the Cubic Galileon simulations considered here is different from the nDGP case, which is just an usual $\Lambda$CDM model expansion history. Additionally, the linear $G_{\rm eff}(a)$ effective gravitational constant in the Cubic Galileon have smaller deviations with respect to GR at earlier times, but, at sufficiently late times, it has a large enhancement of almost two times the GR value. In Figure~\ref{fig:M_conv_cugal} the convergence of the iterative solutions of $M_{\rm it}\left(k, z\right)$ for the Cubic Galileon is shown, and, once again, we see that the curves for $3$ and $5$ iterations are already in great agreement.
%%%%%%%%%%%%%%%%
%%%%%%%%%%%%%%%%
In Figure~\ref{fig:cugal_GR_ratio} we show the agreement between our COLA simulations with respect to the full {\it N}-body simulations of \cite{Barreira:2013eea} with specifications given in the right of Table~\ref{tab:NB_specs}. We can see that our results are able to get an agreement of below $1\%$ at the three redshifts we have {\it N}-body data until $k\approx 3~h/$Mpc. In Appendix~\ref{sec:AppB} we show the comparison between the absolute non-linear power spectra directly between COLA and {\it N}-body, where we see that the force resolution of COLA simulations plays an important role, while the boost remains largely insensitive to this variation. 

%%%%%%%%%%%%%%%%
%%%%%%%%%%%%%%%%

\section{Conclusion}\label{sec:concl}

In this work we have revisited the formalism to introduce screening in modified gravity {\it N}-body simulations proposed by R. Scoccimarro in~\cite{Scoccimarro:2009eu}. We further extended it in Section~\ref{sec:method_mcht} to Horndeski theories characterized by Equation~(\ref{eq:MCHT}), which can be described either by an action, as in the case of the Cubic Galileon model, or by using the EFT of DE approach. We have further implemented this approach in {\it N}-body simulations by using the publicly available \texttt{COLA-FML} code, which implements the COLA method. Our results are presented in Section~\ref{sec:results_ndgp} for the nDGP gravity model, a well-known case in the literature, and in Section~\ref{sec:results_cugal} for the Cubic Galileon model. We have compared the predictions for the ratio of the non-linear matter power spectrum of these MG theories to their GR counterparts from our COLA simulations, which implement this approximate screening method, with full {\it N}-body simulations, which solve the exact equation of motion for the scalar field perturbations. Overall we found an excellent agreement in the scales of interest of future stage-IV LSS surveys, i.e. $k\approx 3h/$Mpc at all redshifts we have {\it N}-body results to compare with.

The formalism of combining the COLA method with Scoccimarro's screening prescription was shown to be an excellent alternative to create new emulators, or to extend current emulators to accommodate beyond-$\Lambda$CDM models, in this case modified gravity. The procedure outlined and validated in the present work also allows us to test theories of gravity in a model-independent fashion, as we have derived all necessary quantities for the implementation of this screening method using the notations and conventions already adopted by the EFT of DE. At the same time, we have computed much of the modified gravity quantities, for nDGP and Cubic Galileon, using the publicly available Einstein-Boltzmann solver \hiclass, while the non-trivial integrals that have appeared in this work were also performed with two publicly available \texttt{python} codes, as pointed out in the text. Therefore, we believe we have outlined a straightforward and reproducible procedure that can be used to test gravity at the non-linear scales, where much of the information coming from future LSS surveys will reside.

The method of including the Vainshtein screening in the effective Newton constant $G_\mathrm{eff}(z,k)$ can be combined with the {\it N}-body gauge methods develoepd in \cite{Brando:2020ouk, Brando:2021jga}. In this method, relativistic corrections are included as linear density fields in simulations so that the power spectrum from {\it N}-body simulations agree with the one computed by the Boltzmann code, e.g. \hiclass, on large scales. The effects of modified gravity on small scales on the other hand are included in the effective Newton constant in {\it N}-body simulations, e.g. $G_\mathrm{eff}(z)$ given by Equation~(\ref{eq:Geff_horn}) for Horndeski theories. The screening can be naturally incorporated by replacing this by $G_\mathrm{eff}(z)$ computed in this paper. We will then be able to model the power spectrum consistently from the largest scales where relativistic effects are important to small scales where screening is important. 

\acknowledgments
For the purpose of open access, the authors have applied a Creative Commons Attribution (CC BY) licence to any Author Accepted Manuscript version arising from this work. We are grateful to Alexandre Barreira and Baojiu Li for kindly sharing their data on the Cubic Galileon simulations used for our validation analysis. We also thank Obinna Umeh for helpful discussions regarding some of the integrals performed in this work. GB is supported by the Alexander von Humboldt Foundation. KK is supported by the UK STFC grant ST/S000550/1 and ST/W001225/1. HAW thanks the Research Council of Norway for their support. Numerical computations were done on the Sciama High Performance Compute (HPC) cluster which is supported by the ICG, SEPNet and the University of Portsmouth. 

\paragraph{Data availability} Supporting research data are available on reasonable request from the corresponding author.

%%%%%%%%%%%%%%
\appendix 

\section{Convergence tests} \label{sec:AppA}

\begin{figure}[h] 
\centering
\includegraphics[width=1.\textwidth]{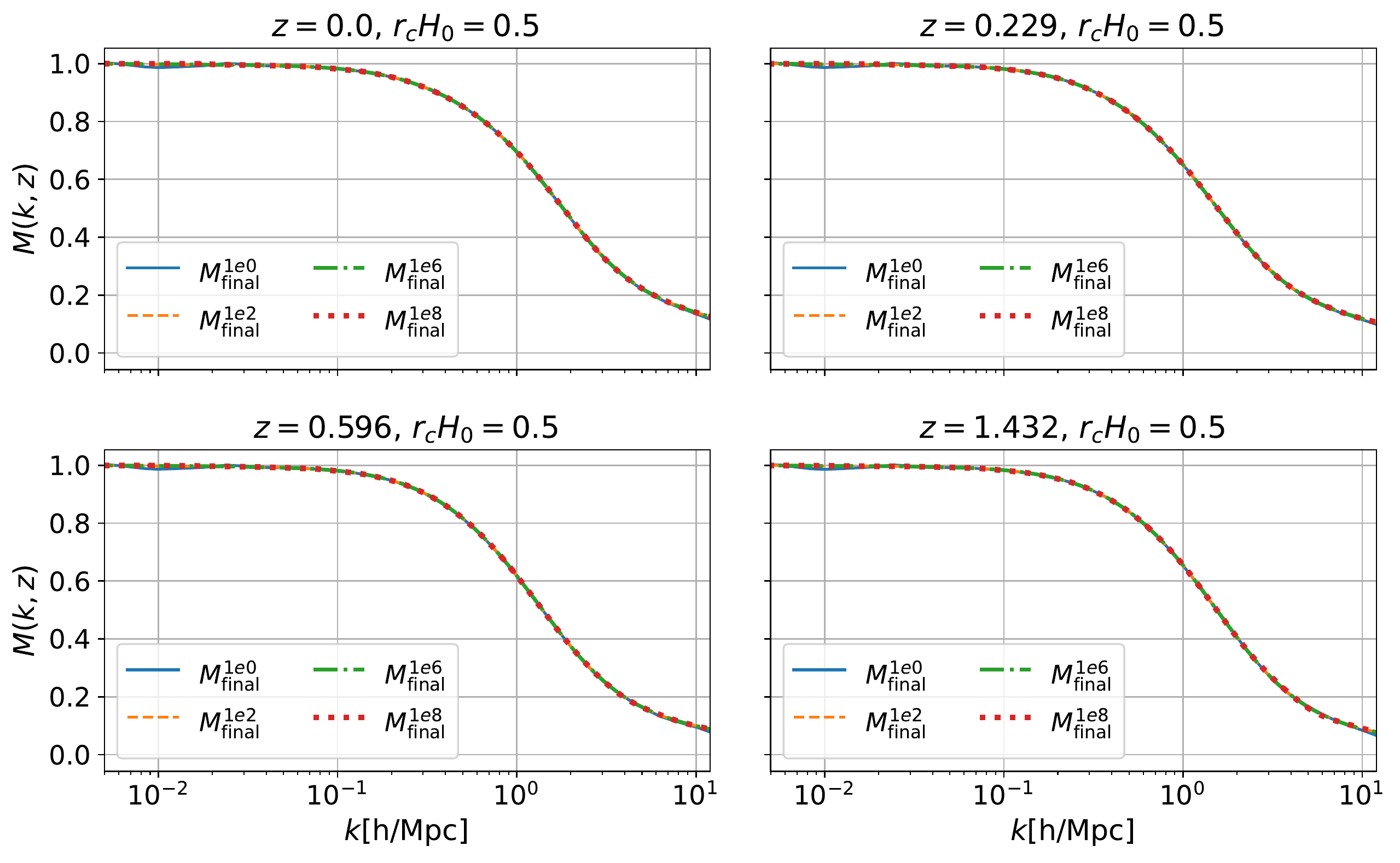}
\caption{Varying the amplitude of $\Delta^{2}\left(k,z\right)$ for the initial guess in our iterative scheme, $M_{\rm ini}\left(k,z\right)$. Each superscript, $1\rm{e}0$, $1\rm{e}2$, $1\rm{e}6$, $1\rm{e}8$, means that we have multiplied $\Delta^{2}\left(k,z\right)$ by: $10^{0}$, $10^{2}$, $10^{6}$ and $10^{8}$ respectively, where the blue curve is the original quantity $\Delta\left(k,z\right)$.}
\label{fig:M_conv_vary_ini}
\end{figure}

In this Appendix we discuss the independence of the final solution for the function $M_{\rm final}\left(k,z\right)$, with respect to the amplitude of the initial guess $M_{\rm ini}\left(k,z\right)$, Equation~(\ref{eq:M_ini}). In order to test this we have chosen to multiply the amplitude of the non-linear dimensionless power spectrum computed using \texttt{halofit} by different orders of magnitude, that is, we have used as our initial guess the following quantity:
\begin{equation}
    \Delta^{2}_{\rm vary}\left(k,z\right) = A \times \Delta^{2}_{\rm orig}\left(k,z\right),
\end{equation}
where $\Delta^{2}_{\rm orig}\left(k,z\right)$ is the original non-linear dimensionless power spectrum, and $A=10^{0}$, $10^{2}$, $10^{6}$ and $10^{8}$. If the iteration process is robust enough, the choice of initial guess should not matter, and the same final solution should always be achieved after some iterations at all redshifts. In Figure~\ref{fig:M_conv_vary_ini} we show the final solution $M_{\rm final}\left(k,z\right)$ for the four values of the constant $A$ mentioned before, at four different redshifts for the most extreme case of nDGP gravity, i.e., $r_{c}H_{0}=0.5$.
%%%%%%%%%%%%%%%%
%%%%%%%%%%%%%%%%
We can see that the different curves, depicted in the legend of each plot, all overlap throughout the considered scales, as well as at different redshifts of the simulation time-steps, which shows that using the same number of iteration as in the main text, $5$ iterations, the same final solution is achieved.

\section{COLA tests} \label{sec:AppB}

In this Appendix we present the results of some additional tests we have performed.

%The COLA simulations used in this Appendix differ only from the ones used in the rest of this paper with respect to how the initial conditions were generated.

We performed COLA simulations with the same initial conditions as the full {\it N}-body simulations to be able to make a direct comparison. However, for these {\it N}-body simulations the only data we have available are the positions and velocities of the particles and because of this, as we already discussed in Section~\ref{sec:sims}, we read particle positions directly from the snapshot at $z=49$ used in the {\it N}-body simulations for Cubic Galileons, and then try to reconstruct the displacement field (which are needed to run COLA simulations) using the Zel'dovich approximation. 

For the Cubic Galileon model, we ran COLA simulations with two different force resolutions, one with the mesh grid per dimension being two times the number of particles per dimension, $N_{\rm mesh}^{\rm 1d} = 2 \times N_{\rm p}^{\rm 1d}$, and a higher resolution one, $N_{\rm mesh}^{\rm 1d} = 3 \times N_{\rm p}^{\rm 1d}$.

\begin{figure}[h] 
\centering
\includegraphics[width=0.85\textwidth]{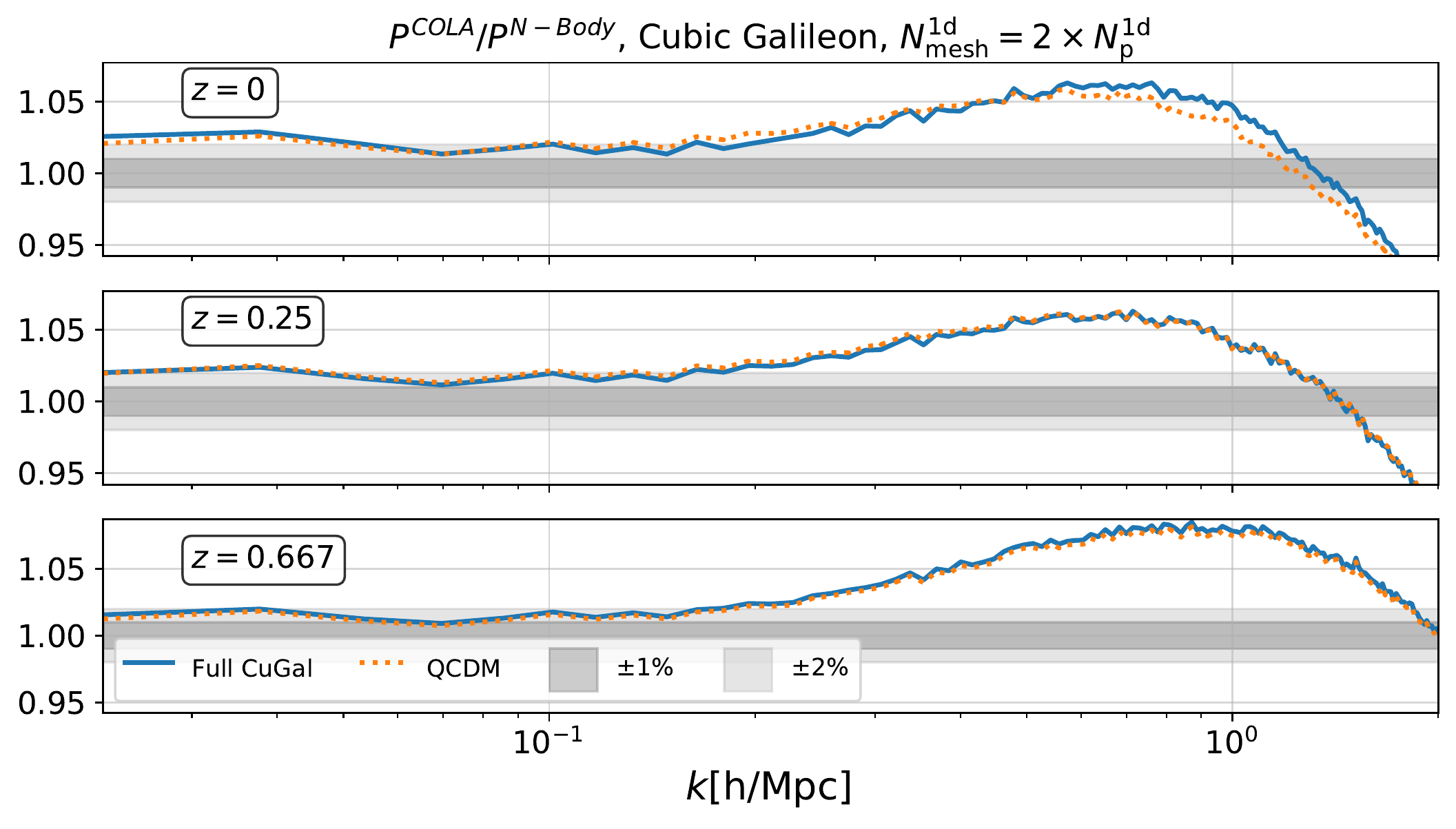}
\caption{Absolute matter power spectrum comparison between COLA and {\it N}-body simulations for the Cubic Galileon Model at three different redshfits for the lower force resolution COLA simulations.}
\label{fig:direct_pk_lowres}
\end{figure}

\begin{figure}[h] 
\centering
\includegraphics[width=0.85\textwidth]{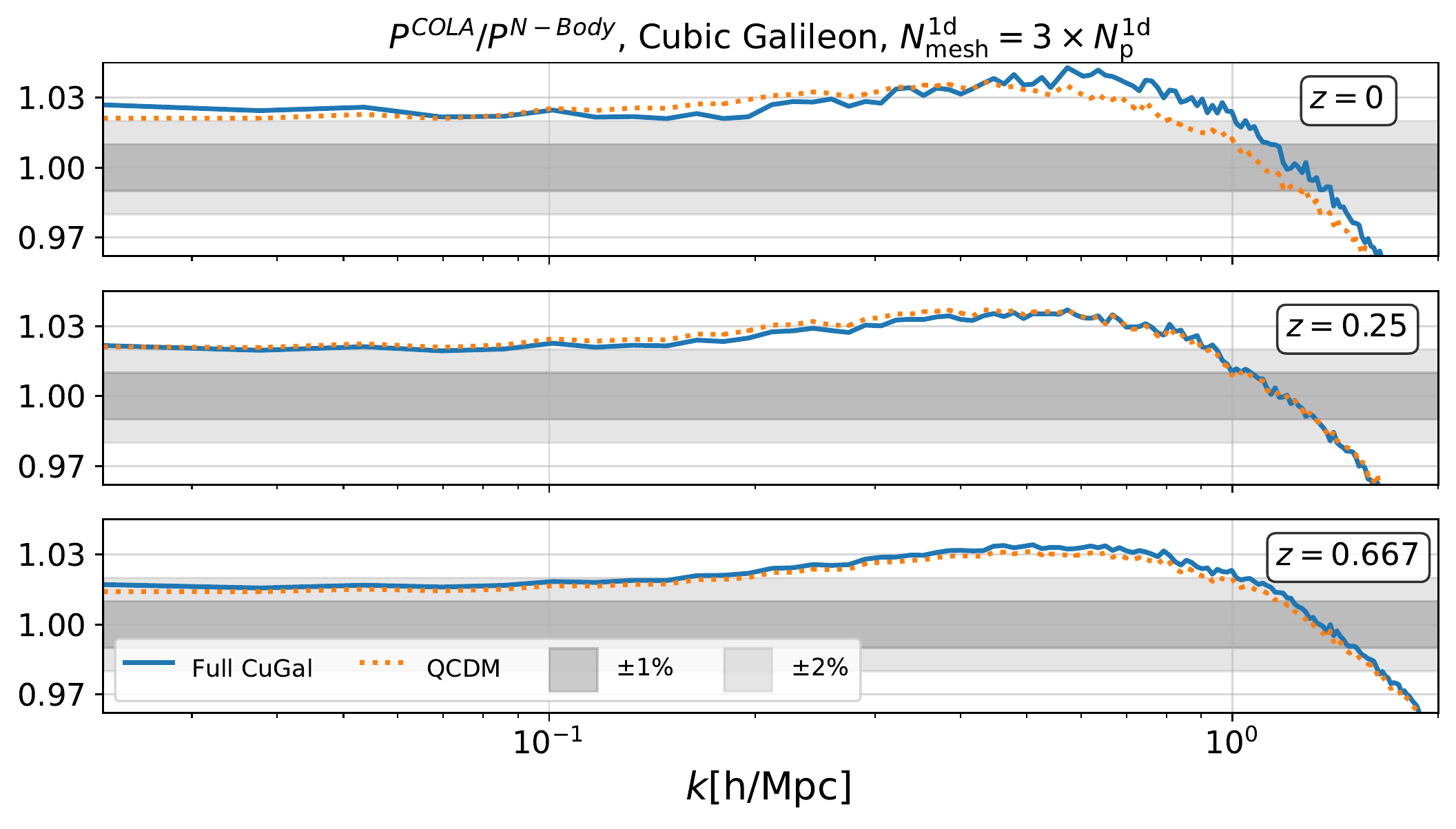}
\caption{Absolute matter power spectrum comparison between COLA and {\it N}-body simulations for the Cubic Galileon Model at three different redshfits for the higher force resolution COLA simulations.}
\label{fig:direct_pk_hires}
\end{figure}

\begin{figure}[h] 
\centering
\includegraphics[width=1.\textwidth]{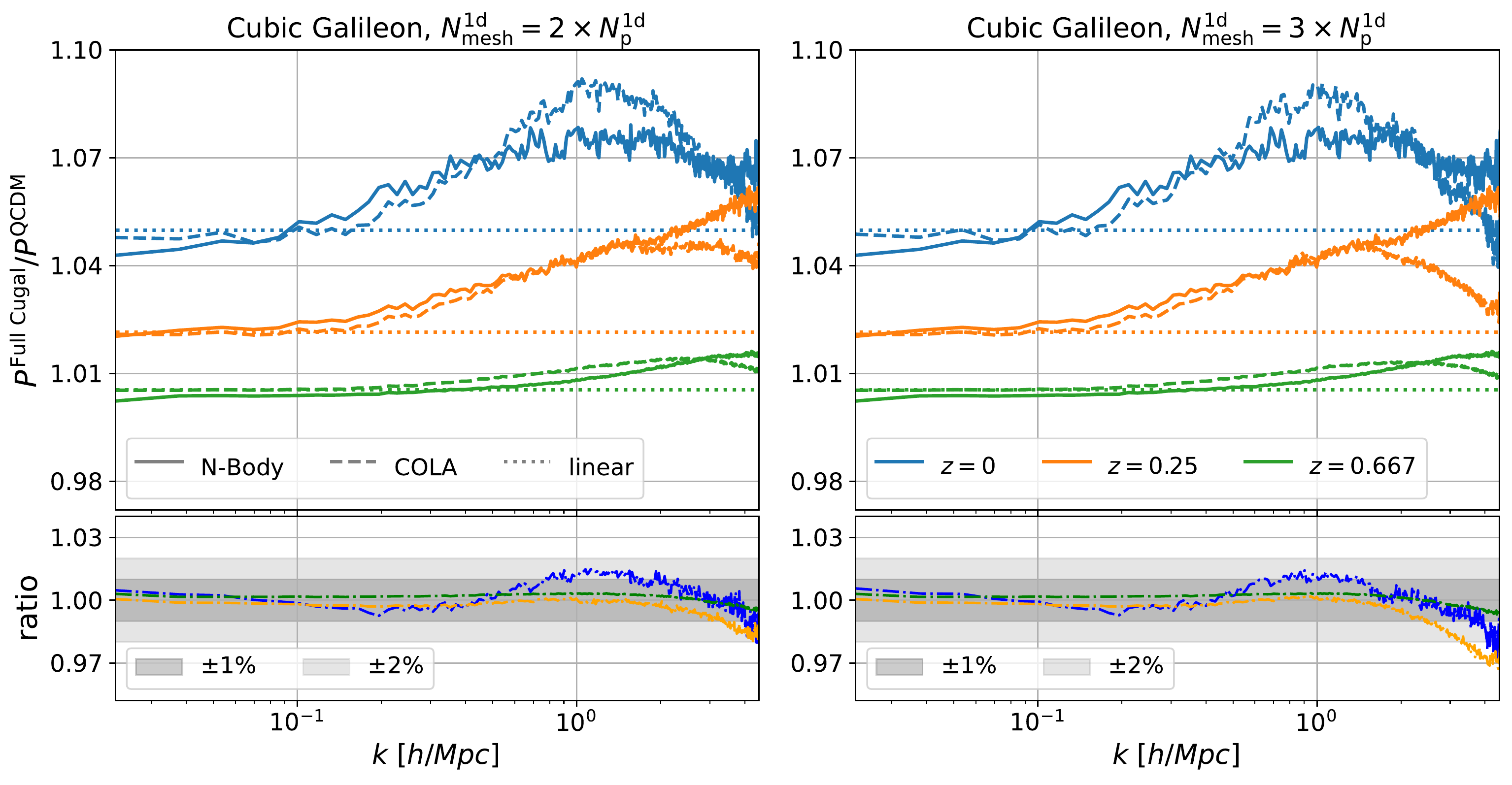}
\caption{Ratio of the power spectrum between the Cubic Galileon model and GR from {\it N}-body simulations (solid lines), COLA (dashed lines) and linear prediction, at three different redshifts, for two force resolution for the COLA simulations: $N_{\rm mesh}^{\rm 1d}=2\times N_{\rm p}^{\rm 1d}$ (left) and $N_{\rm mesh}^{\rm 1d}=3\times N_{\rm p}^{\rm 1d}$ (right) .}
\label{fig:ratio_cugal_GR_boost}
\end{figure}

The comparison of the absolute non-linear matter power spectrum computed using {\it N}-body simulations and these two COLA simulations are quite sensitive to the COLA force resolution, as shown in Figures~\ref{fig:direct_pk_lowres} and~\ref{fig:direct_pk_hires} for the Cubic Galileon model, where there is a clear gain in the agreement between COLA and {\it N}-body when increasing the number of mesh grids. However, it is easily seen that there is a constant offset between the amplitude of the two prescriptions, which is due to the fact that we are not able to reproduce the same expansion history as the one used in the {\it N}-body simulations. The boost function, nevertheless, is generally insensitive to the force resolution of the simulations, as we can see in Figure~\ref{fig:ratio_cugal_GR_boost}, where the same agreement of below $1\%$ at the three redshifts we have {\it N}-body data until $k\approx 3~h/$Mpc is found.

%\guilherme{ The results for the nDGP gravity boost comparison with these generation of ICs is showcased in Figure~\ref{fig:ndgp_GR_ratio_ic}, where we kept the force resolution the same in both panels of the Figure, $N_{\rm mesh}^{\rm 1d}=2\times N_{\rm p}^{\rm 1d}$. Contrary to the Cubic Galileon case, however, in this comparison we are not able to reproduce the particle positions from the {\it N}-body simulation snapshot at $z=49$ exactly. This is an expected difficulty, as recreating the displacement field exactly from a snapshot is known to be a very noisy procedure, specially at high redshifts.}
%the absolute comparison of the non-linear matter power spectrum computed using {\it N}-body simulations and two COLA simulations, one with a smaller force resolution with the number of cell grids per dimension being two times the number of particles per dimension, and one with higher force resolution, number of cell grids being three times the number of particles in Figures~\ref{fig:direct_pk_lowres} and \ref{fig:direct_pk_hires}, respectively.
%%%%%%%%%%%%%%%%
%%%%%%%%%%%%%%%%
%%%%%%%%%%%%%%%%
%%%%%%%%%%%%%%%%
%Contrary to the boost comparisons showcased in Figure~\ref{fig:cugal_GR_ratio}, the power spectrum computed with COLA is sensitive to the force resolution, with the higher configuration settings performing better in reproducing the scale dependence seen in the {\it N}-body simulation until $k\approx 1h/$Mpc. However, it is easily seen that there is a constant offset between the amplitude of the two prescriptions, which is due to the fact that we are not able to reproduce the same expansion history as the one used in the {\it N}-body simulations. Our COLA simulations set 2LPT initial conditions after reading particle information from initial snapshots used in {\it N}-body simulations, while {\it N}-body simulations used Zel'dovich initial conditions to run simulations. This difference creates transient effects at intermediate redshifts as the amplitude of the power spectrum with the Zel'dovich initial condition is suppressed compared with the one with the 2LPT initial condition on non-linear scales. \guilherme{These transient effects, however, are mostly cancelled out when we compare the ratio between non-linear power spectra in the Cubic Galileon case and its GR counterpart (QCDM), as shown in Figure~\ref{fig:ratio_cugal_GR_boost}.}
%%%%%%%%%%%%%%%%

\section{2LPT in minimally coupled Horndeski theories}\label{sec:AppC} 

In this Appendix we show the derivation of the second-order Lagrangian equations of motion for the theories specified by Equation~(\ref{eq:MCHT}). The starting point will be to write Equations~(\ref{eq:Vx_mcht}) and (\ref{eq:poi_mcht}) as:
\begin{subequations}
\begin{align}
    &\partial^{2}V_{X} = \frac{2}{\bra H} \frac{\delta G}{G} \frac{V_{X}^{(2)}}{a^{2}} - \frac{2}{\bra H}\frac{\delta G}{G} \frac{\kappa}{2}a^{2}\rho_{\rm m} \delta,\label{eq:scf_eom_lpt} \\
    &\partial^{2}\Psi = \frac{\kappa}{2}a^{2}\rho_{\rm m} \delta - \frac{\bra H}{2}\partial^{2}V_{X},\label{eq:poi_eom_lpt}
\end{align}
\end{subequations}
where we have used Equation~(\ref{eq:Geff_nl_mcht}). Now, we rewrite Equations~(\ref{eq:scf_eom_lpt}-\ref{eq:poi_eom_lpt}) using the scale factor, $a$, as the time variable:
\begin{subequations}
\begin{align}
    &\partial^{2}v_{X} = \frac{2}{\bra H} \frac{\delta G}{G} \frac{v_{X}^{(2)}}{a^{4}} - \frac{2}{\bra H}\frac{\delta G}{G} \frac{\kappa}{2}a^{4}\rho_{\rm m} \delta, \label{eq:2lpt_2_a}\\
    &a^{2}\partial^{2}\Psi = \frac{\kappa}{2}a^{4}\rho_{\rm m} \delta - \frac{\bra H}{2}\partial^{2}v_{X}\label{eq:2lpt_2_b},
\end{align}
\end{subequations}
where we have introduced:
\begin{equation}
    v_{X} \equiv \frac{\delta \phi}{\phi^{\prime}} = a^{2}V_{X}, \ \ ^{\prime}  \equiv \frac{\dd}{\dd a}.
\end{equation}
Combining Equations~(\ref{eq:2lpt_2_a}) and (\ref{eq:2lpt_2_b}) we arrive at:
\begin{align}\label{eq:2lpt_3}
    a^{2} \partial^{2} \Psi = - \frac{\delta G}{G} \frac{v_{X}^{(2)}}{a^{4}} + G_{\rm eff}\frac{3H^{2}a^{4}\Omega_{\rm m}}{2}\delta
\end{align}
with
\begin{equation}
    \frac{\kappa}{2}\rho_{\rm m} = \frac{3H^{2}\Omega_{\rm m}}{2}.
\end{equation}
From now on we will use $\Omega_{\rm m}(a) = \Omega_{\rm m}$ as the time-evolving fractional energy density of matter, while $\Omega_{\rm m0}$, with the subscript ``$0$", refers to its present value. The two are related as:
\begin{equation}
    \Omega_{\rm m} = \frac{\Omega_{\rm m0}H_{0}^{2}}{H^{2}a^{3}}.
\end{equation}
To make the notation even more compact, we will further define the following quantity:
\begin{equation}
    \Bar{\kappa} = \frac{3H^{2}a^{4}\Omega_{\rm m}}{2} = \frac{3}{2}\Omega_{\rm m0}H_{0}^{2}a,
\end{equation}
then we recast Equation~(\ref{eq:2lpt_3}) as:
\begin{align}\label{eq:poi_eq_lpt}
    a^{2} \partial^{2}\Psi = - \deltaG \frac{v_{X}^{(2)}}{a^{4}} + \barkap \Geff \delta.
\end{align}
Lagrangian perturbation theory (LPT) is given in terms of the mapping:
\begin{equation}\label{eq:lagrang_map}
    \vecx = \vecq + \vec{S}\left( \vec{q}, t\right),
\end{equation}
where the Eulerian particle position, $\vec{x}$, and the Lagrangian particle position, $\vec{q}$, are connected via the displacement field, $\vec{S}\left( \vec{q}, t\right)$. Now, in Eulerian space, the geodesic equation reads:
\begin{equation}
    \frac{\dd^{2} \xxi}{\dd t^{2}} + 2H \frac{\dd \xxi}{\dd t} = - \partial_{i}^{x}\Psi\left(\vecq,t\right),
\end{equation}
with $\partial^{x}_{i} \equiv \partial/\partial x_{i}$. We can rewrite this equation as
\begin{equation}\label{eq:geo_superT}
    \frac{1}{a^{2}}\frac{\dd^{2} \xxi}{\dd T^{2}} = - \partial^{x}_{i} \Psi\left(\vecx,T\right),
\end{equation}
where the super-conformal time variable, $T$, is defined via:
\begin{equation}
    \dd t \equiv a^{2} \dd T.
\end{equation}
Then taking the divergence of Equation~(\ref{eq:geo_superT}) one finds:\begin{equation}\label{eq:geo_superT_div}
    \frac{\dd^{2} \xxii}{\dd T^{2}} = - a^{2}\partial_{i}^{x}\partial_{i}^{x}\Psi\left(\vecx,T\right).
\end{equation}
From the divergence of the Lagrangian mapping, Equation~(\ref{eq:lagrang_map}), we can connect the Eulerian coordinate with the displacement field and write the geodesic equation as:
\begin{equation}
    \frac{\dd^{2}}{\dd T^{2}}\left[\partial^{x}_{i}S_{i}\left(\vecq,T\right) \right]= - a^{2}\partial_{i}^{x}\partial_{i}^{x}\Psi\left(\vecx,T\right).
\end{equation}
The spatial derivatives in the previous equations, however, are still in Eulerian space, and we need to recast them in terms of the Lagrangian coordinates $\vec{q}$, which can be done through the conservation equation:
\begin{align}
    \rho(\vec{x}, T) \dd^{3}x &= \rho(\vec{q}, T) \dd^{3} q,\\
    \rho(\vec{x}, T) &= \left(1 + \delta(\vec{x}, T)\right)\rho(\vec{q}, T),
\end{align}
which gives us 
\begin{equation}\label{eq:inv_jacob}
    J^{-1}(\vec{q},T) \equiv \Bigg|\frac{\partial \vec{q}}{\partial\vec{x}}\Bigg| = 1 + \delta(\vec{x}, T),
\end{equation}
where $J$ is the determinant of the Jacobian of the transformation from $\vec{q}$ to $\vec{x}$.
%where the Jacobian is:
%\begin{equation}
%    \rho(\vec{x}, T) \dd^{3}x = \rho(\vec{q}, T) J\left(\vec{q}, T\right)\dd^{3}q,
%\end{equation}
%and
%\begin{equation}\label{eq:inv_jacob}
%    \frac{1}{J}(\vec{q},T) = \Bigg|\frac{\dd^{3} q}{\dd x^{3}}\Bigg| = 1 + \delta\left( \vec{x}, t \right).
%\end{equation}
From Equation~(\ref{eq:lagrang_map}) we find:
\begin{equation}
    J(\vec{q},T) = \mathrm{det} \left( \delta_{ij} + S_{i,j}(\vec{q},T) \right),
\end{equation}
with
\begin{equation}
    S_{i,j} = \frac{\partial S_{i}}{\partial q_{j}},
\end{equation}
where partial derivatives without a superscript are to be taken with respect to the Lagrangian position $\vecq$, and from now on we will omit the explicit spatial and time dependence on the functions which are already in Lagrangian space. The chain rule gives us:
\begin{align}\label{eq:chain_rule}
    \partial_{i}^{x} = \frac{\dd^{3}q_{j}}{\dd x^{3}_{i}} \partial_{j}^{q} = \frac{1}{\delta_{ij} + S_{i,j}}\partial_{j}^{q},
\end{align}
and, we can use the following approximation to rewrite Equation~(\ref{eq:chain_rule})
\begin{equation}
    \frac{1}{\delta_{ij} + S_{i,j}} \approx \delta_{ij} - S_{i,j} + \dots.
\end{equation}
Equation~(\ref{eq:geo_superT}) then becomes:
\begin{equation}\label{eq:geo_superT_q}
    \frac{\dd^{2} \Sii}{\dd T^{2}} - \Sij \frac{\dd^{2} \Sji}{\dd T^{2}} = - a^{2}\partial^{2}_{x}\Psi(\vecx, T),
\end{equation}
and using Equation~(\ref{eq:poi_eq_lpt}) we can write Equation~(\ref{eq:geo_superT_q}) as:
\begin{equation}\label{eq:geo_superT_lpt}
    \frac{\dd^{2} \Sii}{\dd T^{2}} - \Sij \frac{\dd^{2} \Sji}{\dd T^{2}} = \deltaG \frac{v_{X}^{(2)}(\vecx, T)}{a^{4}} - \barkap \Geff \delta(\vecx, T).
\end{equation}
We now need to recast the right-hand-side of this equation in Lagrangian space, and, in order to do so, we begin by expanding the fields up to second order:
\begin{subequations}
\begin{align}
    &\delta = \epsilon \delta^{(1)} + \epsilon^{2} \delta^{(2)} + \dots, \label{eq:delta_exp}\\
    &S = \epsilon S^{(1)} + \epsilon^{2} S^{(2)} + \dots,\\
    &v_{X} = \epsilon \hat{v}_{X}^{(1)} + \epsilon^{2} \hat{v}_{X}^{(2)} + \dots.
\end{align}
\end{subequations}
By inspecting Equation~(\ref{eq:geo_superT_lpt}), the scalar field fluctuations appear already at second order, therefore, we define:
\begin{equation}
    v_{X}^{(2)} \equiv \hat{v}_{X}^{(1,2)} = \left(\partial^{2}\hat{v}_{X}^{(1)}\right)^{2} - \partial_{i}\partial_{j}\partial^{i}\partial^{j}\hat{v}^{(1)}_{X}.
\end{equation}
Linearizing Equation~(\ref{eq:scf_eom_lpt}), we find the following expression
\begin{equation}\label{eq:vx_lpt1}
    \partial^{2}\hat{v}_{X}^{(1)}(\vecx, T) = - \frac{2}{\bra H} \deltaG \barkap\delta^{(1)}(\vecx, T).
\end{equation}
From Equation~(\ref{eq:inv_jacob}) we can write the Jacobian as a function of the displacement field:
\begin{align}
    \frac{1}{J(\vec{q},t)} = 1 + \delta(\vec{x}, t), \ \ J = |\delta_{ij} + S_{i,j}(\vec{q}, t)|,
\end{align}
then we can show that\footnote{Where we have used the following identities:
\begin{align*}
    &\mathrm{det}\left(I+A\right) \approx I + \mathrm{tr}(A) + \frac{1}{2}\left[ \mathrm{tr}^{2}(A) - \mathrm{tr}(A^{2}) \right] \dots,\\
    &\frac{1}{\mathrm{det}\left(I+A\right)} \approx I - \mathrm{tr}(A) + \frac{1}{2}\left[ \mathrm{tr}^{2}(A) + \mathrm{tr}(A^{2}) \right] \dots.
\end{align*}}:
\begin{subequations}
\begin{align}\label{eq:jacob_exp}
    &J \approx 1 + S_{i,i}^{(1)} + S_{i,i}^{(2)} + \frac{1}{2}\left[ \left(S_{i,i}^{(1)}\right)^{2} -  S_{i,j}^{(1)}S_{i,j}^{(1)}\right],\\
    &J^{-1} \approx 1 - S_{i,i}^{(1)} - S_{i,i}^{(2)} + \frac{1}{2}\left[ \left(S_{i,i}^{(1)}\right)^{2} +  S_{i,j}^{(1)}S_{i,j}^{(1)}\right],
\end{align}
\end{subequations}
and from Equation~(\ref{eq:delta_exp}) we get: 
\begin{subequations}
\begin{align}
    &\delta^{(1)}(\vec{x}, t) = - S_{i,i}^{(1)}\label{eq:delta_lpt1}, \\
    &\delta^{(2)}(\vec{x}, t) = - S_{i,i}^{(2)}  + \frac{1}{2}\left[ \left(S_{i,i}^{(1)}\right)^{2} +  S_{i,j}^{(1)}S_{i,j}^{(1)}\right]\label{eq:delta_lpt2}.
\end{align}
\end{subequations}
Due to the nature of the expansion of the Jacobian up to second order Equation~(\ref{eq:jacob_exp}), we further rewrite the displacement field $\vec{S}(\vecq, T)$ as the gradient of a scalar field, $S = \nabla \phi$, and, in Fourier space, this is simply
\begin{subequations}
\begin{align}\label{eq:S_exp_lpt}
    &\mathcal{F}\left[S_{i}(\vec{q}, T)\right]\left(\vec{k}\right) = i k_{i}\phi(\vec{k},T), \\
    &\mathcal{F}\left[S_{i,i}(\vec{q}, T)\right]\left(\vec{k}\right) = -k^{2}\phi(\vec{k},T).
\end{align}
\end{subequations}
At linear order, we can combine Equations~(\ref{eq:vx_lpt1}) and (\ref{eq:delta_lpt1}) to find
\begin{align}
    \frac{\dd^{2} S^{(1)}_{i,i}}{\dd T^2} = -\barkap \Geff \delta^{(1)} = \barkap \Geff S_{i,i}^{(1)},
\end{align}
which can then be written in terms of the linear gradient field:
\begin{align}
    -k^{2}\frac{\dd^{2} \phi^{(1)}}{\dd T^{2}}\left(\vec{k}, T\right) = -k^{2} \barkap \Geff \phi(\vec{k},T).
\end{align}
This equation is separable, so we can further separate the solution as:
\begin{equation}
    \phi\left(\veck, T\right) = D_{1}(T)\phi\left(\veck, T_{0}\right),
\end{equation}
where $\phi\left(\veck, T_{0}\right)$ is the initial first order scalar field displacement, and $D_{1}(T)$ is the usual first order growth factor, that can be found by solving
\begin{equation}
    \left[\frac{\dd^{2}}{\dd T^{2}} - \barkap \Geff\right]D_{1}(T) = 0.
\end{equation}

Following the same procedure as above, we arrive at the equation of motion for the second order displacement field:
\begin{align}
    \frac{\dd^{2}S_{i,i}^{(2)}}{\dd T^{2}} - S_{i,j}^{(1)}\frac{\dd^{2} S_{j,i}^{(2)}}{\dd T^{2}} = \deltaG \frac{\hat{v}^{(1,2)}}{a^{4}} - \barkap \Geff \left\lbrace -S_{i,i}^{(2)} + \frac{1}{2}\left[ \left( S_{i,i}^{(1)} \right)^{2} + S_{i,j}^{(1)}S_{j,i}^{(1)} \right] \right\rbrace.
\end{align}
Using Equations~(\ref{eq:vx_lpt1}), (\ref{eq:delta_lpt1}) and (\ref{eq:delta_lpt2}) we find
\begin{align}\label{eq:eq_2lpt}
    \left[\frac{\dd^{2}}{\dd T^{2}} - \barkap \Geff\right]S_{i,i}^{(2)}  =\left[\frac{\barkap^{2}}{a^{4}} \frac{4}{\bra^{2}H^{2}}\left(\deltaG\right)^{3}-\frac{1}{2}\barkap \Geff\right]\left[\left(S_{i,i}^{(1)}\right)^{2} - S_{i,j}^{(1)}S_{j,i}^{(1)}\right].
\end{align}
Transforming to Fourier space gives us:
\begin{align}\label{eq:2lpt_pre_final}
    \left[\frac{\dd^{2}}{\dd T^{2}} - \barkap \Geff\right]\mathcal{F}\left[S_{i,i}^{(2)}\left(\vecq, T\right)\right]\left(\veck, T\right) = \left[\frac{\barkap^{2}}{a^{4}} \frac{4}{\bra^{2}H^{2}}\left(\deltaG\right)^{3}-\frac{1}{2}\barkap \Geff\right]D_{1}^{2}(T) \times \mathcal{I}\left(\veck, T\right),
\end{align}
where
\begin{equation}
    \mathcal{I}\left(\veck, T\right) = \int \frac{\dd^{3}k_{1}\dd^{3}k_{2}}{\left(2\pi\right)^{3}}\delta^{\rm D}\left(\veck - \veck_{12}\right)\left[1-\left(\vec{\hat{k}}_{1}\cdot \vec{\hat{k}}_{2}\right)^{2}\right] \delta_{0}^{(1)}\left(\veck_{1}\right)\delta_{0}^{(1)}\left(\veck_{2}\right).
\end{equation}
We can further rewrite this expression as:
\begin{align}
    \left[\frac{\dd^{2}}{\dd T^{2}} - \barkap \Geff\right]\left(-k^{2}\right)\phi^{(2)}\left(\vec{k}, T \right) = -\frac{\Geff \barkap}{2}\left[ 1 - \frac{4 \barkap^{2}}{a^{4}\bra^{2}H^{2}} \left(\deltaG\right)^{3} \frac{2}{\Geff \barkap} \right] D_{1}^{2}(T) \times \mathcal{I}\left(\vec{k}, T \right),
\end{align}
and if we define:
\begin{align}\label{eq:phi_2lpt_1}
    \phi^{(2)} \equiv -\frac{1}{2k^{2}} \int \frac{\dd^{3}k_{1}\dd^{3}k_{2}}{\left(2 \pi\right)^{3}} \delta^{\rm D}\left( \vec{k} - \vec{k}_{12} \right)D_{2}\left(\vec{k}, \veck_{1}, \veck_{2}, T\right) \delta_{0}^{(1)}\left(\veck_{1}\right)\delta_{0}^{(1)}\left(\veck_{2}\right),
\end{align}
we arrive at an expression for the second order growth factor $D_{2}$:
\begin{equation}\label{eq:D2_LPT}
    \left[\frac{\dd^{2}}{\dd T^{2}} - \barkap \Geff\right]D_{2}\left(\vec{k}, \veck_{1}, \veck_{2}, T\right) = - \barkap \Geff D_{1}^{2}(T)\left[ 1 - \frac{2a^{4}H^{2}}{\barkap \Geff}f(a) \right]\left[1-\left(\vec{\hat{k}}_{1}\cdot \vec{\hat{k}}_{2}\right)^{2}\right],
\end{equation}
where
\begin{equation}
    f(a) = \frac{H_{0}^{2}}{H^{2}} ~ \frac{H_{0}^{2}}{\bra^{2}H^{2}} ~ \frac{9\Omega_{\rm m0}^{2}}{2a^{6}}\left(\deltaG\right)^{3}.
\end{equation}
We can compare this expression to the already well-known case in the literature for nDGP, Equation~(5.9) of~\cite{Winther:2017jof}:
\begin{equation}
    f(a) = \frac{H_{0}^{2}}{H^{2}}~ r_{c}^{2}H_{0}^{2}~ \frac{9\Omega_{\rm m0}^{2}}{2a^{6}}\left(\deltaG\right)^{3}.
\end{equation}
Similarly to the linear case, Equation~(\ref{eq:D2_LPT}) is also separable, as the operator acting on $D_{2}$ on the left-hand side of this equation is only time-dependent, which allows us to rewrite Equation~(\ref{eq:phi_2lpt_1}) as:
\begin{align}\label{eq:phi_2lpt_2}
    \phi^{(2)} \equiv -\frac{1}{2k^{2}} D_{2}(T) \int \frac{\dd^{3}k_{1}\dd^{3}k_{2}}{\left(2 \pi\right)^{3}} \delta^{\rm D}\left( \vec{k} - \vec{k}_{12} \right) \left[1-\left(\vec{\hat{k}}_{1}\cdot \vec{\hat{k}}_{2}\right)^{2}\right]\delta_{0}^{(1)}\left(\veck_{1}\right)\delta_{0}^{(1)}\left(\veck_{2}\right).
\end{align}
In this way Equation~(\ref{eq:D2_LPT}) simply becomes:
\begin{equation}\label{eq:D2_LPT_2}
    \left[\frac{\dd^{2}}{\dd T^{2}} - \barkap \Geff\right]D_{2}\left(T\right) = - \barkap \Geff D_{1}^{2}(T)\left[ 1 - \frac{2a^{4}H^{2}}{\barkap \Geff}f(a) \right].
\end{equation}
%%%%%%%%%%%%%%%%%%%%%%%%%%%%%%%%% 

\bibliographystyle{JHEP}
\bibliography{bib_my}

\end{document}